\documentclass{pasa}
\usepackage{wrapfig, subcaption, booktabs, longtable, caption, makecell}
\usepackage{gensymb}
\usepackage{threeparttable}
\usepackage{graphicx}
\usepackage{qtree}
\usepackage[linguistics]{forest}

\usepackage{tikz}
\usetikzlibrary{shapes,arrows}
\usepackage{booktabs}

\usepackage[usestackEOL]{stackengine}[2013-10-15]
\def\x{\hspace{3ex}}    
\stackMath

\title[Imaging at 300\,MHz With the MWA]{A Calibration and Imaging Strategy at 300\,MHz With the Murchison Widefield Array (MWA)}

\author[J. H. Cook et al.]{J. H. Cook$^{1}$, N. Seymour$^{1}$, and M. Sokolowski$^{1}$
\affil{$^1$International Centre for Radio Astronomy Research, Curtin University, Perth, Australia}%
}%

\jid{PASA}
\doi{10.1017/pas.\the\year.xxx}
\jyear{\the\year}

\usepackage{aas_macros}
\usepackage{hyperref} 
\hypersetup{colorlinks,citecolor=blue,linkcolor=blue,urlcolor=blue}

\begin{document}

\begin{frontmatter}
\maketitle

\begin{abstract}
 At relatively high frequencies, highly sensitive grating sidelobes occur in the primary beam patterns of low frequency aperture arrays (LFAA) such as the Murchison Widefield Array (MWA). This occurs when the observing wavelength becomes comparable to the dipole separation for LFAA tiles, which for the MWA occurs at $\sim300$\,MHz. The presence of these grating sidelobes has made calibration and image processing for 300\,MHz MWA observations difficult. This work presents a new calibration and imaging strategy which employs existing techniques to process two example 300\,MHz MWA observations. Observations are initially calibrated using a new 300\,MHz sky-model which has been interpolated from low frequency and high frequency all-sky surveys. Using this 300\,MHz model in conjunction with the accurate MWA tile primary beam model, we perform sky-model calibration for the two example observations. After initial calibration a self-calibration loop is performed by all-sky imaging each observation with \textsc{wsclean}. Using the output all-sky image we mask the main lobe of the image. Using this masked image we perform a sky-subtraction by estimating the masked image visibilities using \textsc{wslcean}. We then image the main lobe of the observations with \textsc{wsclean}. This results in high dynamic range images of the two example observation main lobes. These images have a resolution of $2.4$ arcminutes, with a maximum sensitivity of $\sim31\,\textrm{mJy/beam}$. The calibration and imaging strategy demonstrated in this work, opens the door to performing science at 300\,MHz with the MWA, which was previously an inaccessible domain. With this paper we release the code described below and the cross-matched catalogue along with the code to produce a sky-model in the range $70-1400$\,MHz.
\end{abstract}

\end{frontmatter}

\section{Introduction}\label{Sec:Introduction}

The Murchison Widefield Array (MWA) is a low frequency aperture array (LFAA), and is a precursor to the Square Kilometre Array LFAA (SKA\_LOW). The MWA is an interferometer which has several primary radio science goals: observation of the epoch of reionisation by detecting the statistical cosmologically redshifted 21cm neutral hydrogen signal; Galactic and extra-galactic science, through observing synchrotron and HII radio emissions in the Galaxy, and in extra-galactic radio galaxies; observing transient objects such as pulsars and fast radio bursts; studying the ionosphere, heliosphere, and solar activity \citep{MWA2,MWA_sci,MWA_sci-2}. The MWA is comprised of 256 tile stations capable of observing in the frequency range of $70-320\,\textrm{MHz}$, with an instantaneous bandwidth of $\Delta\nu = 30.72\,\textrm{MHz}$ \citep{MWA2,MWA-Cor,MWAPH2}. Each tile is a collection of 16 dual-polarised dipoles arranged in a $4\times4$ North-South, East-West grid.

The number and layout of the MWA tiles provides an excellent snapshot $uv$-coverage, in conjunction with the typical widefield of view for LFAAs, the MWA is well suited to quickly surveying the entire sky. As part of the the Galactic and extra-galactic all-sky MWA survey \citep[GLEAM;][]{GLEAM}, the MWA observed the entire radio sky south of declination (DEC) $+25\degree$ in the frequency range $72-231\,\textrm{MHz}$. The MWA additionally observed the entire sky at $300$\,MHz during an extended observing run in the second year of the GLEAM survey.

Unlike dish arrays the MWA electronically points by introducing a delay in the signal between each dipole in a tile. The tile then combines each dipole to form a beam response pattern on the sky (the primary beam) \citep{MWA1}. For most of the MWA frequency range the primary beam is dominated by the main lobe which is aligned with the pointing direction. However, at high frequencies when the dipole separation is comparable to the observing wavelength, highly sensitive sidelobes known as gratings lobes appear in the primary beam pattern. These grating lobes are reflections of the main lobe and appear above the horizon at a frequency around $\sim300$\,MHz for the MWA \citep{BEAM2014}. At the higher frequency MWA regime ($\nu \gtrsim 280\,\textrm{MHz}$), the MWA can be more sensitive to radio emission in the grating lobes than the main lobe. Bright sources present in these grating lobes introduce point spread function (PSF) sidelobe structures that affect the sensitivity of the main lobe. The higher frequency regime is also heavily affected by RFI, with the lower 280\,MHz band of the MWA having a high RFI occupancy \citep{RFI}. As a result of RFI, observations are typically limited to the $70-230$\,MHz frequency range. These issues make calibrating and imaging 300\,MHz MWA observations more complex than at lower frequencies, and until now the high frequency regime has largely been neglected.

The frequency dependence of LFAA observations, means at higher frequencies the MWA has better resolution. At 300\,MHz the Phase II MWA extended configuration \citep{MWAPH2} has a resolution comparable to the $45\,\textrm{arcsec}$ resolution of the 1.4 GHz NRAO VLA Sky Survey \citep[NVSS;][]{NVSS}. This allows for a direct comparison between dish arrays and LFAAs, leading to a better understanding of the systematic differences between the different kinds of radio interferometers. Observations at 300\,MHz will also further constrain the spectral energy densities (SEDs) of radio sources; along with higher resolutions at 300\,MHz this will aid in the classification of radio sources. Sources with peaks at higher frequencies may also be too faint to detect at lower MWA frequencies.

Due to the squared wavelength $(\lambda^2)$ dependence of Faraday rotation, polarised radio sources at higher MWA frequencies are significantly less depolarised than sources at lower MWA frequencies \citep{WST-pol}. Grating lobes also pose a problem for polarisation studies, therefore a calibration method for 300\,MHz observations could open the door to more sensitive MWA polarisation observations. Calibrated 300\,MHz polarisation observations would add high value to existing low frequency polarisation work \citep{Lenc-MWA-pol,POGS,POGS2}.

With a good understanding of the MWA tile beam model and a model of the sky at 300\,MHz, it is possible to calibrate MWA observations in the high frequency MWA regime. In this work we describe a 300\,MHz sky-model catalogue which is constructed from cross-matched low and high frequency catalogues. In particular this work uses GLEAM and NVSS to interpolate the sky flux density at 300\,MHz. We also use the fully embedded element \citep[FEE;][]{BEAM2016} MWA tile beam model. The FEE MWA tile beam models each coarse channel in the frequency range $72-315\,\textrm{MHz}$. Using the FEE beam model in conjunction with the 300\,MHz sky-model, we can calibrate MWA observations using the sky-model calibration method. In particular we use the direction independent calibration software \textsc{calibrate} \citep{calibrate}. \textsc{calibrate} is based on the direction independent part of the \textsc{Mitchcal} algorithm \citep{Mitchcal}, which uses an apparent sky-model generated by the MWA tile beam and a sky catalogue to calibrate the gain amplitude and phases for each tile. In this work we use \textsc{calibrate} to process a calibrator 300\,MHz MWA observation. The calibration solutions from this observation can then be transferred to another observation at the same pointing taken on the same observing run.

This paper is divided into the following sections: Section \ref{Sec:Catalogues} discusses the low and high frequency catalogues that are interpolated to create the 300\,MHz sky-model; Section \ref{Sec:Sky-model} details the sky-model, and the SED fitting processes; Section \ref{Sec:Observations} introduces the observations used to demonstrate the calibration and imaging strategy used in this work; Section \ref{Sec:Calibration} discusses the calibration strategy; Section \ref{Sec:Imaging} discusses the imaging strategy; Section \ref{Sec:Results} introduces the images produced from the test observations; Section \ref{Sec:Discussion-Conlcusion} discusses the results and concludes the work.

\section{Low \& High Frequency Catalogues}\label{Sec:Catalogues}

In this section we describe the low and high frequency catalogues that cover the sky below $\textrm{DEC} \leq +45\,\degree$, in the frequency range $72\,\rm{MHz}$ to $1.4\,\rm{GHz}$. The majority of the sky is covered by the GLEAM extra-galactic catalogue \citep[GLEAM\_exGal;][]{GLEAMyr1}. This catalogue forms the basis for the 300\,MHz sky-model discussed further in (Section \ref{Sec:Sky-model}). GLEAM\_exGal is missing several regions, particularly around the Galactic plane (GP), the large and small Magellanic clouds (LMC, SMC), around Centaurus A (CenA), and in two wedge shaped regions. GLEAM\_exGal is also missing the exceptionally bright `A-team' radio sources, as well regions above $\textrm{DEC} \geq +30\,\degree$. These missing regions can be filled in with later releases of the GLEAM data \citep{GLEAM-LMC-SMC,GLEAM-GP}, as well as from other high and low frequency surveys such as NVSS and the TIFR Giant Metrewave Radio Telescope (GMRT) $150\,\rm{MHz}$ all sky radio survey ADR1 \citep[TGSS;][]{TGSS}. These surveys are discussed in Sections \ref{Sec:GLEAM_Sup} and \ref{Sec:TGSS/NVSS}. Additionally we describe several bespoke A-team source models which were created from existing GLEAM images, and supplemented with high accuracy spectral energy density (SED) models from \citet{Perl}. These A-team source models are critical for calibration purposes and are discussed in Section \ref{Sec:A-team}.

\subsection{PUMA Catalogue}\label{Sec:PUMAcat}

Properly cross-matching catalogues with different sensitivities and resolutions is complex, therefore we use the Positional Update and Matching Algorithm \citep[PUMA]{PUMA}. We use a PUMA created catalogue which combines the GLEAM\_exGal catalogue with higher frequency catalogues (J. Line, personal communications). We hereon refer to the PUMA catalogue as PUMAcat. PUMAcat was created by cross-matching the GLEAM\_exGal catalogue with the following list of surveys: the 74 MHz Very Large Array Low Frequency Sky Survey redux \citep[VLSSr;][]{VLSSr}; TGSS \citep{TGSS}; the 843 MHz Sydney University Molonglo Sky Survey \citep[SUMSS;][]{SUMSS}; NVSS \citep{NVSS}. 

PUMAcat contains $308,584$ radio sources and covers a frequency range of $72\,\rm{MHz}$ to $1.4\,\rm{GHz}$, including (where possible) the full GLEAM bands from $72\,\rm{MHz}$ to $231\,\rm{MHz}$. Table \ref{table:PUMA-table} breaks down the different source types in PUMAcat. The sources classified \texttt{isolated}, \texttt{multiple}, and \texttt{dominant} in Table \ref{table:PUMA-table} are defined in \citet{PUMA}. The $5,203$ sources defined as \texttt{N/A} are GLEAM\_exGal sources which did not have any corresponding matches in the other catalogues. The $783$ \texttt{Aegean} sources are a bespoke extended source model developed by \citet{PUMA} for the EoR0\footnote{Epoch of Reionisation Field 0 (EoR0) centred at $\textrm{RA} = 0^h$ and $\textrm{DEC} = -27^d$.} field.

\begin{table}[t]
\centering
\begin{tabular}{@{}ll@{}}
\toprule
 Match Type & N  \\ \midrule
 \texttt{isolated} & $257,583$  \\
 \texttt{multiple} & $33,783$   \\
 \texttt{dominant} & $4,753$  \\
 \texttt{N/A}& $5,203$ \\
 \texttt{Pietro} & $6,460$ \\ 
 \texttt{Aegean} & $783$ \\\bottomrule
\end{tabular}
\caption[]
{\small Break down of the different match type sources in PUMAcat.}
\label{table:PUMA-table}
\end{table}

The last class of sources are the $6,460$ \texttt{Pietro} sources from \citet{Proc} which were derived from a deep six hour MWA survey of the EoR1\footnote{Epoch of Reionisation Field 1 (EoR1) centred at $\textrm{RA} = 4^h$ and $\textrm{DEC} = -30^d$.} field in the 182\,MHz band. In PUMAcat these sources have their own spectral values for the frequencies $170$\,MHz, $190$\,MHz and $210$\,MHz. These values come from fitting across the $182$\,MHz band with a second order polylogarithmic function. In \citet{Proc} this was done to ensure smooth spectral behaviour when calibrating their observations. As a result these sources do not have any quoted errors in PUMAcat. To estimate the error in the \texttt{Pietro} bands we calculated the median relative error for each of the GLEAM subbands in PUMAcat. We then fit the relative error in the subbands with a second order polynomial. Using the second order fit we estimated the relative error in each of the fitted \texttt{Pietro} bands and updated PUMAcat. We additionally filtered $20$ sources from PUMAcat which either had one or no flux density measurements.

\subsection{GLEAM Supplementary Sky-Model}\label{Sec:GLEAM_Sup}

The aforementioned missing regions in GLEAM\_exGal can be filled in using the GLEAM 200\,MHz sky-model created to calibrate MWA Phase II data. This sky-model is constructed from processed publicly available GLEAM data \citep{GLEAM}. Specifically it includes missing regions from recent publications \citep{GLEAM-LMC-SMC,GLEAM-GP}, and unpublished processed public GLEAM data around CenA \citep{GLEAM-X}. The main purpose of this GLEAM sky-model is to process the GLEAM extended survey (GLEAM-X) data (Hurley-Walker et al. in prep). This model is publicly available through the \href{https://github.com/nhurleywalker/GLEAM-X-pipeline}{GLEAM-X}\footnote{https://github.com/nhurleywalker/GLEAM-X-pipeline} GitHub repository \citep{GLEAM-X}. The GLEAM sky-model additionally contains multi-component A-team source models for Hydra A and Virgo A, these will be further discussed in Section \ref{Sec:A-team}.

For the purposes of this work we only require a subset of the GLEAM sky-model which covers the missing regions in PUMAcat. We cross-matched the GLEAM sky-model to PUMAcat at a separation of $2$\,arcminutes. The GLEAM sky-model sources that were found to not have matches with PUMAcat were formed into a subset catalogue. Excluding the two wedge regions, this subset catalogue contains $48,816$ sources which cover the missing GP, LMC, SMC and CenA regions in GLEAM\_exGal. This subset catalogue is hereon referred to as the GLEAM supplementary catalogue (GLEAM\_Sup).

\begin{figure*}[t]
            \centering 
            \includegraphics[width=\textwidth]{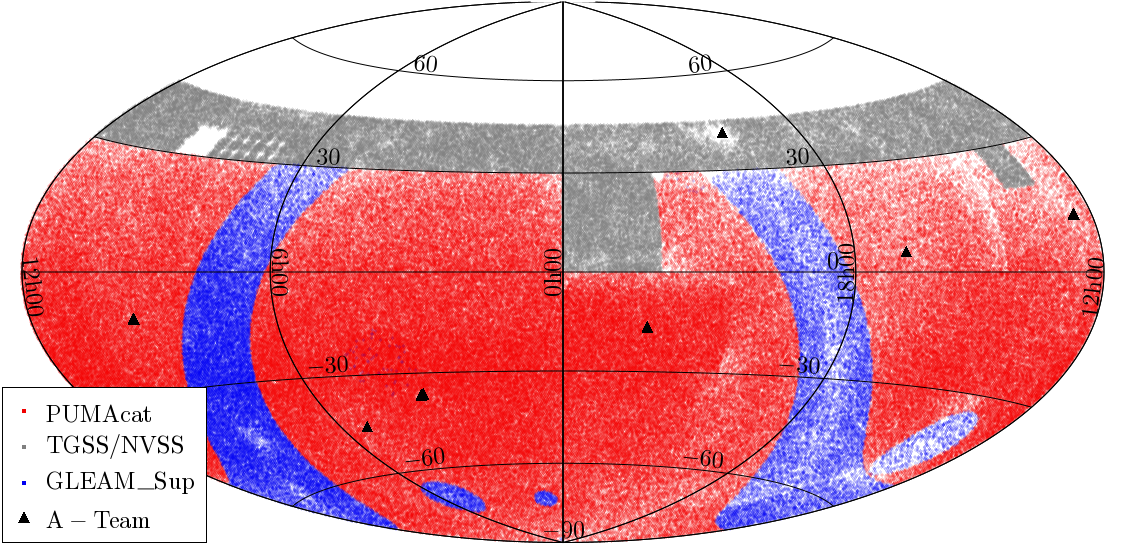}
            \caption{Aitoff projection showing the sky coverage of PUMAcat (red), TGSS/NVSS (light grey) subset and the GLEAM\_Sup catalogue (blue). The black triangles indicate the position of A-Team sources. Gaps are present in the TGSS/NVSS catalogue at $\textrm{DEC} \geq +30\degree$, these gaps are a result of missing data in the TGSS catalogue \citep{TGSS}. Additional gaps occur at the boundary between the TGSS/NVSS catalogue, and the other catalogues.}
            \label{fig:Total-sky-map}
\end{figure*}

\subsection{TGSS/NVSS Spectral Index Catalogue}\label{Sec:TGSS/NVSS}

The remaining regions which need to be filled in are the two missing wedge regions, and declinations higher than $+30\,\degree$. For this work we use the TGSS/NVSS spectral index catalogue, where \citet{Intema} cross-matched the first TGSS \citep{TGSS} data release with NVSS \citep{NVSS}. This catalogue covers the frequency range $150\,\rm{MHz}$ to $1.4\,\rm{GHz}$, and the entire sky above declination $-40\,\degree$. Importantly this work investigated the spectral index $\alpha^{150}_{1400}$ of these sources. This is useful because it allows for the interpolation of the 300\,MHz flux density. Interpolation will be discussed further in Section \ref{Sec:TGSS/NVSS-flux-density}

In this work we took only the single (S), multiple (M) and Complex (C) sources from the catalogue (for a definition of these sources see \citet{Intema}). Sources with only a single detection in either NVSS or TGSS were ignored since the sensitivity of both of these surveys is deeper than the GLEAM survey. Additionally sources identified as island (I) were internally cross-matched to removed double detections. These are hold overs from the cross-matching method used by \citet{Intema} to join the two catalogues. After the filtering process the total number of sources for each of the three subset regions is given in Table \ref{table:TGSS/NVSS}.

\begin{table}[t]
\centering
\begin{tabular}{@{}llll@{}}
\toprule
 Region & RA Range & DEC Range & $N$ \\ 
 & [deg] & [deg] & \\ \midrule
 Wedge 1 & $[196,209]$  & $[20,30]$ & $1,779$ \\
 Wedge 2 & $[320,360]$ & $[0,30]$ & $13,408$\\
 $\textrm{DEC} \in [30,45]$ & $[0,360]$  & $[30,45]$ & $60,729$\\ \bottomrule
\end{tabular}
\caption[]
{\small Break down of the final number of sources in the two wedge regions, and the declination strip from $+30\degree < \textrm{DEC} \leq +45\degree$.}
\label{table:TGSS/NVSS}
\end{table}

\subsection{A-Team Sources}\label{Sec:A-team}

The A-team sources are a set of exceptionally bright radio galaxies which are not present in GLEAM\_exGal \citep{GLEAMyr1}. Some of these sources are many arcminutes to degrees in size (Fornax A, and Centaurus A for example), others such as Pictor A (PicA) are not fully resolved at lower MWA frequencies, and might only be partially resolved at 300\,MHz. Due to the magnitude of their brightness these sources are often used as calibrators for interferometric radio observations. As such accurate SEDs and spatial models for these sources are necessary for creating accurate calibration solutions for 300\,MHz MWA observations. 
 
Some multi-component Gaussian models exist in the GLEAM sky-model for Hydra A and Virgo A. For the remaining A-team sources we created bespoke point source models for for PicA, 3C444, Cygnus A, Hercules A and Fornax A using cutout GLEAM images\footnote{\href{http://gleam-vo.icrar.org/gleam_postage/q/form}{http://gleam-vo.icrar.org/gleam\_postage/q/form}} \citep{GLEAM}. The highest resolution 227\,MHz GLEAM band cut out images for these sources were used. For the unresolved sources only a single point source model was fit. For partially resolved sources we used NVSS cutout images\footnote{\href{https://www.cv.nrao.edu/nvss/postage.shtml}{https://www.cv.nrao.edu/nvss/postage.shtml}}, and fit two point sources since these sources might be resolved at 300\,MHz \citep{NVSS}. Spatially resolved complex sources such as Fornax A required more attention. We fit 18 point sources to Fornax A, one to the core, and 17 to the lobes. Point source regions were specifically placed at prominent bright features in the lobes of Fornax A. Due to its complex nature, Fornax A is not typically used as a calibrator source, as such a highly accurate model was not the focus of the science in this work. The SEDs for the A-team sources will be discussed in Section \ref{Sec:A-Team SED's}

\subsection{Low \& High Frequency Catalogue Sky Coverage}\label{Sec:Sky-coverage}

The sky coverage of the aforementioned catalogues extends across the entire sky below $\textrm{DEC} < +45\degree$, and can be seen in Figure \ref{fig:Total-sky-map}. There are notable gaps in the sky coverage, particularly in the right ascension (RA) range $97\degree.5 \leq \textrm{RA} \leq 142\degree.5$ and DEC range $25\degree\leq \textrm{DEC} \leq 39\degree$. This region is missing from the first data release of TGSS. These missing regions come from a set of 10 TGSS observations with particularly bad ionospheric conditions \citep{TGSS}. In addition to this missing region, gaps between the GLEAM and TGSS/NVSS based catalogues are visible at their boundaries in Figure \ref{fig:Total-sky-map}. These gaps along with fewer sources found in bright regions such as the GP and CenA, will affect the completeness of the total sky coverage in these regions, but are not detrimental to this work.

\section{300\,MHz Sky-Model}\label{Sec:Sky-model}

\subsection{PUMAcat 300\,MHz SED Models}\label{Sec:PUMAcat-SEDmodels}

\begin{figure*}[t]
    \begin{center}
        \begin{subfigure}[b]{0.495\textwidth}
            \centering
            \includegraphics[width=\textwidth]{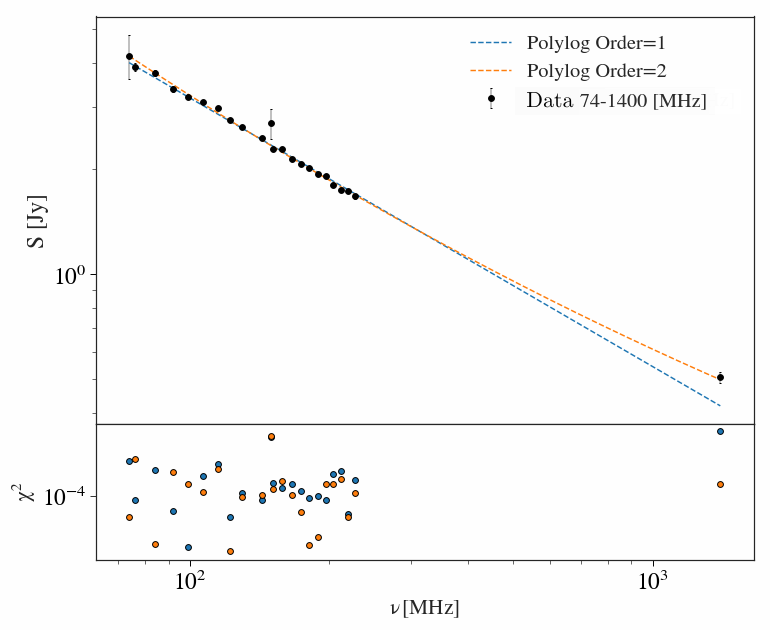}
            \caption[]%
            {{\small Preferential power-law Fit}}
            \label{fig:power-law-ex}
        \end{subfigure}
        \hfill
        \begin{subfigure}[b]{0.495\textwidth}
            \centering 
            \includegraphics[width=\textwidth]{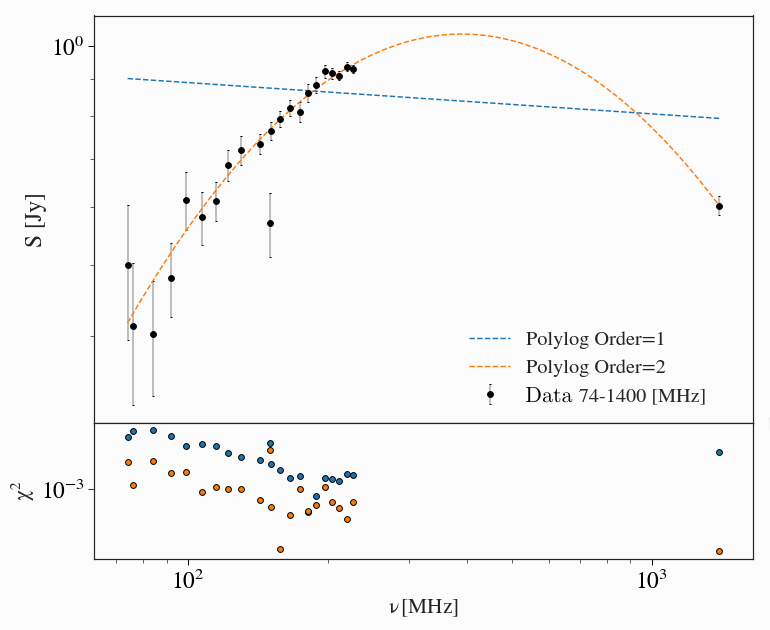}
            \caption[]%
            {{\small Preferential Second Order Polylogarithmic Fit}}
            \label{fig:2nd-poly-ex}
        \end{subfigure}
        \caption
        {Log-log plot of the SED of representative sources taken from PUMAcat. In the top panel the black circles are the normalised flux densities as a function of frequency. The dashed blue line is the power-law fit to the SED (first order polylogarithmic fit), the dashed orange line is the second order polylogarithmic fit to the SED. The bottom panel shows the $\chi^2$ normalised residuals for both fits as a function of frequency, where the colours correspond to the model in the top panel. Subfigure \ref{fig:power-law-ex} shows a source with a preferred power-law fit, and Subfigure \ref{fig:2nd-poly-ex} shows a source with a preferred second order polylogarithmic fit.}
        \label{fig:model-fits}
    \end{center}
\end{figure*}

In this work we fit two models in log-space to the SEDs of radio sources in PUMAcat. These models allowed for the interpolation of the 300\,MHz flux density. The first model we fit to each radio source was a power-law model\footnote{Also defined as a first order polylogarithmic model.}:

\begin{equation}\label{eq:polylog1}
\log_{10}(S_\nu) = \log_{10}(S_{\nu_0}) + \alpha \left( \log_{10}\left(\frac{\nu}{\nu_{0}}\right) \right)
\end{equation}
where $\nu_{0}$ is the reference frequency which we define at 300\,MHz, $S_{\nu_0}$ is the flux density at the reference frequency, and $\alpha$ is the spectral index. Equation \ref{eq:polylog1} in log-space is a straight line model, where the spectral index $\alpha$ is the gradient, and $\log_{10}(S_{\nu_0})$ is the y-intercept. For radio sources that only had two or three flux density measurements, we only fit the power-law model. The second model we fit is a second order polylogarithmic (polylog) function, this model is simply a parabola defined in log-space:

\begin{multline}\label{eq:polylog2}
    \log_{10}(S_\nu) = \log_{10}(S_{\nu_0}) + \alpha \left( \log_{10}\left(\frac{\nu}{\nu_{0}}\right) \right)  + \\ q \left( \log_{10}\left(\frac{\nu}{\nu_{0}}\right) \right)^2.
\end{multline}

Equation \ref{eq:polylog2} is the second order approximation of the radio source SEDs in log-space, where the parameter $q$ is the curvature term of the parabola. This model is a natural choice, since many radio sources will display some curvature in their spectra across large enough frequency ranges \citep{GPS,Harvey}. The curvature term $q$ provides a good approximation for sources intrinsic spectral curvature. Additionally $q$ has also been linked to the magnetic field strength of active galactic nuclei (AGN) \citep{Bridle}. Radio sources with $q<0$ indicate spectra with concave\footnote{Here we define concave as a downward opening parabola, convex is defined as the opposite.} curvature, and radio sources with $q>0$ indicate spectra with convex curvature. The spectral index $\alpha$ in Equation \ref{eq:polylog2} represents the steepness of the parabola at the reference frequency $\nu_{0}$. Other models such as broken power-law models do exist \citep{GPS}, but Equation \ref{eq:polylog2} is compatible with the calibration software used in this work, and adequately describes the SED in nearly all curved cases. The significance of this will be discussed further in Section \ref{Sec:Calibration}.
 
The optimal fit parameters for Equations \ref{eq:polylog1} and \ref{eq:polylog2}, are determined by minimising the $\chi^2$ value for each fit. The $\chi^2$ is defined below:

\begin{equation}\label{eq:chisqd}
    \chi^2 = \sum^n_{i=0} \frac{\left(\log_{10}\left(S(\Vec{\theta}|\nu_i)\right) - \log_{10}\left(S_{\textrm{data,i}}\right)\right)^2}{\sigma_i^2},
\end{equation}

where $S(\Vec{\theta}|\nu_i)$ is the model at $\nu_i$ and $S_{\textrm{data},i}$ is the measured flux density at $\nu_i$. In Equation \ref{eq:chisqd}, $\Vec{\theta}$ is a vector which contains the model fit parameters, and $\sigma_i$ is the uncertainty for $\log_{10}\left(S_{\textrm{data},i}\right)$. To perform the fit we use the \textsc{numpy} function \textsc{polyfit} in \textsc{Python}.
 
Model discernment between Equation \ref{eq:polylog1} and \ref{eq:polylog2} is performed by calculating the Bayesian information criterion \citep[BIC]{BIC}:

\begin{equation}\label{eq:BIC}
    \textrm{BIC} = \chi^2 + \ln{(n)}k.
\end{equation}

The \textrm{BIC} takes into consideration the fit to the model $\chi^2$, the number of data points $n$, and the number of model fit parameters $k$. The term $\ln{(n)}k$ penalises models with large numbers of parameters $k$. Models with too many parameters can overfit the data. For each radio source in PUMAcat, we calculate the $\textrm{BIC}$ for both models. We then compute the absolute difference $(\Delta\textrm{BIC})$ between the two models, this difference provides a relative comparison between the two fits. Values of $\Delta \textrm{BIC} \geq 6$ provide strong evidence that the model with the lower \textrm{BIC}, is the preferred fit to the data \citep{BIC-significance}. When $\Delta \rm{BIC} < 6$ there is no significant evidence to select one model over the other. In this case the default preferred fit is Equation \ref{eq:polylog1} since it has fewer fit parameters $k$.
 
Figure \ref{fig:model-fits} illustrates two example SEDs. Subfigure \ref{fig:power-law-ex} shows a source where Equation \ref{eq:polylog1} is a preferred fit. Subfigure \ref{fig:2nd-poly-ex} shows a radio source with a concave SED, with Equation \ref{eq:polylog2} being a clearly preferential fit. Once the sources were fit, and a preferential model was selected, we interpolated the 300\,MHz flux density. The resulting sky-model at 300\,MHz is hereon referred to as PUMA300.

\subsection{GLEAM\_Sup 300\,MHz Model}\label{Sec:GLEAM300-SED's}

Radio sources in the GLEAM\_Sup sky-model were fit using Equations \ref{eq:polylog1} and \ref{eq:polylog2}, at a reference frequency of $\nu_0 = 200\,\textrm{MHz}$ (Hurley-Walker et al., in prep). Since the GLEAM\_Sup sky-model does not contain high frequency information, we extrapolated the flux density at 300\,MHz using the model fit parameters. We did this by transforming the fit coefficients from a reference frequency of $\nu_0 = 200\,\textrm{MHz}$ to $\nu_0 = 300\,\textrm{MHz}$. The full fit coefficient transformation process can be found in Section \ref{sec:poly-log-proof} of the Appendix. For sources with second order polylogarithmic fits, the spectral index $\alpha_{200}$ is now defined at $\alpha_{300}$, where the subscript indicates the reference frequency in MHz. 

\subsection{TGSS/NVSS 300\,MHz Model}\label{Sec:TGSS/NVSS-flux-density}

We interpolated the 300\,MHz flux density for the TGSS/NVSS300 catalogue using a power-law:

\begin{equation}\label{eq:TGSS/NVSS flux}
    S_{300} = S_{\rm{NVSS}}\left( \frac{300}{1400} \right)^{-\alpha}.
\end{equation}

The spectral index in Equation \ref{eq:TGSS/NVSS flux} was determined by \citet{Intema} from the TGSS/NVSS catalogue. We use this spectral index along with the NVSS flux density $S_{\textrm{NVSS}}$ from the TGSS/NVSS catalogue to interpolate the 300\,MHz flux density.

\begin{figure}[t]
            \centering 
            \includegraphics[width=0.47\textwidth]{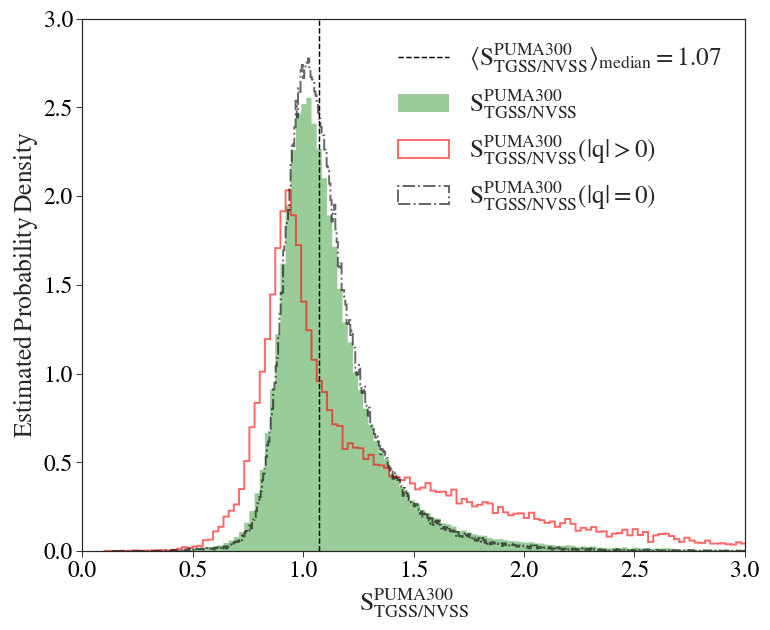}
            \caption{The ratio of the PUMA300 and TGSS/NVSS 300\,MHz flux densities is illustrated by the green histogram. The empty black dot dashed histogram, indicates the PUMA300 sources which were preferentially fit with a power-law $(|q|=0)$. The empty solid red histogram shows the PUMA300 sources with a preferential second order polylogarithmic fit $(|q|>0)$. All histograms show a characteristic skew towards higher ratios, specifically for sources with $|q|>0$. The median flux ratio is shown as the dashed black line.}
            \label{fig:S-ratio-hist}
\end{figure}

We investigated systematic offsets in the interpolated TGSS/NVSS 300\,MHz flux density, relative to the PUMA300 catalogue by performing a cross-match. In this case only single radio sources were considered because they are unresolved in both TGSS/NVSS and PUMA300. Using a cross-match separation of two arcminutes, $176,073$ matches were found. The ratio of the PUMA300 300\,MHz flux density with the TGSS/NVSS estimate was computed, the flux ratio distribution can be seen in Figure \ref{fig:S-ratio-hist}. The median ratio of the distribution is $\langle S^{\textrm{PUMA300}}_{\textrm{TGSS/NVSS}} \rangle = 1.07$ which is close to the expected value of $1$. 

A ratio of $\langle S^{\textrm{PUMA300}}_{\textrm{TGSS/NVSS}} \rangle>1$ indicates an underestimate in the 300\,MHz flux density for the TGSS/NVSS catalogue. There are several potential contributing factors to the underestimate. The primary cause is likely due to the lack of curvature present in the TGSS/NVSS SEDs. In Figure \ref{fig:S-ratio-hist} the distribution is broken into the power-law sources ($q=0$, black dashed line), and sources with curved SEDs ($|q|>0$, red line). The distribution with curved SEDs clearly has a larger tail at $\langle S^{\textrm{PUMA300}}_{\textrm{TGSS/NVSS}} \rangle>1$ when compared to the power-law source flux ratio distribution. The higher sensitivity of the GLEAM\_Sup catalogue (and by extension PUMA300) to extended emission compared to TGSS and NVSS, would also contribute to the underestimate. Additionally there are systematic differences between the PUMA300 flux scale and the TGSS/NVSS flux scale that contribute to the underestimate. To correct for this average underestimate we use the median flux density ratio to scale the TGSS/NVSS300 flux densities.

\subsection{A-Team SED Models}\label{Sec:A-Team SED's}

The total estimated 300\,MHz flux density for each A-team source, was interpolated using SED models fit by \citet{Perl}. These models were developed to investigated the flux scale of A-team sources from $50\,\rm{MHz}$ to $50\,\rm{GHz}$, by fitting arbitrary order polylogarithmic functions at a reference frequency of $\nu_0 = 1\,\rm{GHz}$ \citep{Perl}. Using the coefficient transformation method discussed in the Appendix Section \ref{sec:poly-log-proof}, we transformed the fit coefficients for the A-team sources to a reference frequency of $\nu_0 = 300\,\textrm{MHz}$. The transformed fit coefficients for the total A-team SEDs can be found in Table \ref{table:Perl-ATEAM-1GHz}.

\begin{table*}[t]
\small
\centering
\begin{tabular}{@{}llllll@{}}
\toprule
  Source & $S_{300} = a^{300}_0$ & $\alpha_{300} = a^{300}_1$ & $q_{300} = a^{300}_2$ & $a^{300}_3$ & $a^{300}_4$ \\\midrule
  Fornax A & $2.193\pm0.003$ & $-0.661\pm0.006$ & &  & \\ 
  Pictor A & $2.3144 \pm 0.001$ & $-0.6696\pm0.001$ & $-0.074\pm0.005$ &  & \\ 
  Hydra A & $1.6152\pm0.01$ & $-0.948\pm0.001$ & $-0.0857\pm0.004$ & $-0.0072\pm0.001$ & $0.03\pm0.003$\\
  Virgo A & $2.3021\pm0.0007$ & $-0.8285\pm0.002$ & $-0.048\pm0.003$ & & \\
  Hercules A & $2.3396\pm0.0007$ & $-0.9252\pm0.0009$ & $-0.0951\pm0.0020$ & & \\
  Cygnus A & $3.8122\pm0.0010$ & $-0.7726\pm0.0014$ & $-0.1905\pm0.006$ & $0.0669\pm0.002$ & $0.043\pm0.005$ \\
  3C444 & $1.6229\pm0.0009$ & $-0.9897\pm0.002$ & $0.0458\pm0.004$ & $-0.077\pm0.005$ & \\
 \bottomrule
\end{tabular}
\caption[]
{\small The polylogarithmic coefficients for each of the A-team sources at a reference frequency of $\nu_0 = 300\,\textrm{MHz}$. These were determined by transforming the polylogarithmic coefficients from \citet{Perl} from $\nu_0 = 1000\,\textrm{MHz}$ to $\nu_0 = 300\,\textrm{MHz}$. Columns three and four show the spectral index $\alpha_{300}$ and curvature term $q_{300}$ for each source were applicable. Columns five and six are the higher order polylogarithmic coefficients. These last columns demonstrate the level of curvature present in radio SEDs.}
\label{table:Perl-ATEAM-1GHz}
\end{table*}

Each A-team source contains multiple components, but to simplify the approach we assume that each component has the same SED as the total radio source. We determine the 300\,MHz flux density for each component as a fraction of the total source flux density $S_{\textrm{tot},300}$. This fraction is determined from the bespoke models described in Section \ref{Sec:A-team}, where the flux density of each component is summed at the frequency of the cutout image $S_{\textrm{tot},\nu}$. The fraction for each component is then determined, and multiplied by the total estimated 300\,MHz flux density from the \citet{Perl} SED models.
 
A-team sources are important to consider separately because they are used for calibration purposes \citep{GLEAM,GLEAMyr1}. Particularly in this work, we use PicA to calibrate 300\,MHz MWA observations. These calibration solutions can then be transferred to other observations at the same pointing.

\subsection{Total 300\,MHz Catalogue}\label{Sec:Total300}

The columns for each component catalogue were standardised and concatenated together. The new combined 300\,MHz sky-model catalogue contains $433,345$ table entries, and is hereon referred to as Total300. Table \ref{table:tot-cat-cols} provides the column format for the Total300 sky-model catalogue. Using the method outlined in the Appendix Section \ref{sec:poly-log-proof}, Total300 can be transformed from a 300\,MHz sky-model to one in the frequency range $72-1400\,\rm{MHz}$. The Total300 sky-model and the transformation script are publicly available through the GitHub repository associated with this work \href{https://github.com/JaidenCook/300-MHz-Pipeline-Scripts}{\textsc{S300-pipeline}}\footnote{https://github.com/JaidenCook/300-MHz-Pipeline-Scripts}.
 
Table \ref{table:tot-mod-stats} breaks down the statistics for the component catalogues of Total300 (not including the A-team models). The fraction of sources with $|q_{300}| > 0$ for the combined PUMA300 and GLEAM\_Sup models is $\sim0.163$. The GLEAM\_Sup sources predominantly have positive curvature because they lack the high frequency information.
%
\begin{table}[t]
\centering
\begin{threeparttable}
\centering
\begin{tabular}{@{}llll@{}}
\toprule
  Column Name & Format & Notes \\\midrule
  Name\tnote{a} & - & Unique NVSS source  \\
  & & identification of \\
  & & the format $\rm{JHMS\pm DMS}$\\
  RA & degree & Right ascension\\ 
  DEC & degree & Declination\\
  PA & degree & Position angle\\
  Major & degree & Major axis\\
  Minor & degree & Minor axis\\
  Fint300 & Jy & Total integrated flux\\
  & & density\\
  coefficients\tnote{b} & Tuple & SED polylogarithmic\\
  & & coefficients\\
  Flag & Integer & Subset flag\\
 \bottomrule
\end{tabular}
\begin{tablenotes}\footnotesize
\item[a] The naming convention takes exception to A-team sources where the format of their names is laid out in subsection \ref{Sec:A-team}.
\item[b] The coefficients are formatted in a tuple of size $6$, where $(a_0, a_1,\cdots,a_6)$, sources with only power-law or second order polylogarithmic fits have coefficients $a_3$ to $a_6$ set at 0.
\end{tablenotes}
\caption[]
{\small Column format of the Total300 sky-model catalogue.}
\label{table:tot-cat-cols}
\end{threeparttable}
\end{table}
 
\begin{table*}[t]
\centering
\begin{tabular}{@{}lllll@{}}
\toprule
  Catalogue & $\langle S_{300} \rangle\,[\rm{Jy}]$ & $\langle \alpha_{300} \rangle$ & $\langle q_{300} \rangle$ & $N$ \\\midrule
  PUMA300 & $0.095\pm0.135$ & $-0.814\pm0.253$ & $-0.170\pm0.882$ & $308,563$\\ 
  GLEAM\_Sup & $0.136\pm0.194$ & $-0.799\pm0.604$ & $0.271\pm1.330$ & $48,816$\\ 
  TGSS/NVSS300 & $0.091\pm0.117$ & $-0.736\pm0.275$ & N/A & $75,916$\\
  Total300 & $0.0981\pm0.140$ & $-0.805\pm0.269$ & $-0.098\pm0.908$ & $433,345$\\
 \bottomrule
\end{tabular}
\caption[]
{\small Median values of the SED fits for the three main subsets of the Total300 sky-model. The A-team sources are not included here except for the total number of table entries.}
\label{table:tot-mod-stats}
\end{table*}

\section{Observations}\label{Sec:Observations}

We use publicly available two minute MWA Phase I snapshot observations, which can be downloaded from the MWA All-Sky Virtual Observatory (\href{https://github.com/MWATelescope/manta-ray-client}{ASVO})\footnote{https://asvo.mwatelescope.org/services} server using the MWA \texttt{manta-ray}\footnote{https://github.com/MWATelescope/manta-ray-client} python client \citep{Sokolowski-II}. The raw observation files are downloaded and consolidated into a measurement set using the software \textsc{cotter} \citep{measurement-set,cotter}. \textsc{cotter} averages MWA observation data in time and frequency, it additionally flags radio frequency interference (RFI) using the \textsc{aoflagger} algorithm \citep{aoflagger}. \textsc{aoflagger} was found to be inadequate at flagging most of the RFI at 300\,MHz. In particular there is a high RFI occupancy in the lower coarse channels, as found by \citet{RFI}. As a result the first four coarse channels for every 300\,MHz observation typically have to be flagged. We additionally expanded our flagging regime through the common astronomy software applications (CASA)\footnote{http://casa.nrao.edu/} package \citep{CASA}. In particular we use the CASA RFI flagging functions \textsc{rflag} and \textsc{tfcrop} (refer to \citet{CASA} for a detailed description of these algorithms). Further flagging per baseline is also required, since some baselines in particular have a high occupancy of RFI. To flag these baselines another flagging tool referred to here as \href{https://gitlab.com/Sunmish/piip/blob/master/ms_flag_by_uvdist.py}{\textsc{steflag}}\footnote{https://gitlab.com/Sunmish/piip/blob/master/
ms\_flag\_by\_uvdist.py} is used \citep{Steflag}. This tool flags baseline using their statistics to identify outliers, it then outputs a list of antenna pairs which can be passed to CASA for flagging.

\begin{table}[t]
\centering
\begin{tabular}{@{}lllll@{}}
\toprule
 Name & UTC & GPS Time & RA & DEC \\ 
 &  & $[s]$ & [deg] & [deg]\\ \midrule
 ObsA & 2015-11-08 & 1131042024 & 79.2 & -47.6 \\
 & 18:20:07 & & & \\
 ObsB & 2015-11-08 & 1131038424 & 64.2 & -47.6\\
 & 17:20:07 & & & \\\bottomrule
\end{tabular}
\caption[]
{\small List of example observations used in this work.}
\label{table:Obs}
\end{table}

Table \ref{table:Obs} lists the snapshot 300\,MHz MWA observations used to demonstrate the imaging and calibration strategy outlined in this work. These observations were taken during an extended observing run, in the second year of GLEAM. ObsA is a calibrator observation of the partially resolved A-team radio galaxy PicA. ObsB was taken an hour before ObsA during the same observing night. Both observations have the same pointing. In this work we use ObsA as a calibrator observation to calibrate ObsB. 

\section{300\,MHz Calibration Strategy}\label{Sec:Calibration}

300\,MHz MWA calibrator observations are calibrated using the software package \textsc{calibrate} \citep{calibrate}. This software takes a model of the apparent sky as a function of frequency, and uses this model to predict the visibilities for that observation. \textsc{calibrate} then performs a minimisation with the measured visibilities, to determine the instrumental gain and phase solutions \citep{Mitchcal}. Once derived, the solutions can then be applied to the interferometric data for that observation, or for other observations at the same pointing. The apparent sky-model required to determine the gain amplitude and phase solutions, can be constructed from the Total300 catalogue and the FEE MWA tile beam model \citep{BEAM2016}. Each observation will have a different `apparent sky-model' due to the RA and DEC of the phase centre.

\subsection{Constructing the Apparent Sky Model}

Each snapshot observation has a particular UTC time, this can be used in conjunction with the RA and DEC to determine the zenith $(\theta)$ and a azimuth $(\phi)$ angles for each source in the Total300 catalogue. Sources below the horizon $(\theta > 90\,\degree)$ are removed. The integrated flux density for the remaining sources is then attenuated by the 300\,MHz MWA FEE tile beam response. The brightest 1500 sources are then selected, these sources constitute the base of the apparent sky-model. A model of the apparent flux density $S_\textrm{app}(\nu)$ for the remaining 1500 sources, is required by \textsc{calibrate} in order to predict the observation visibilities:

\begin{equation}\label{eq:Sapp-mod}
    S_{\textrm{app}}(\nu) = B_{\theta,\phi}(\nu) \cdot S(\nu).
\end{equation}

\begin{figure}[t]
            \centering 
            \includegraphics[width=0.48\textwidth]{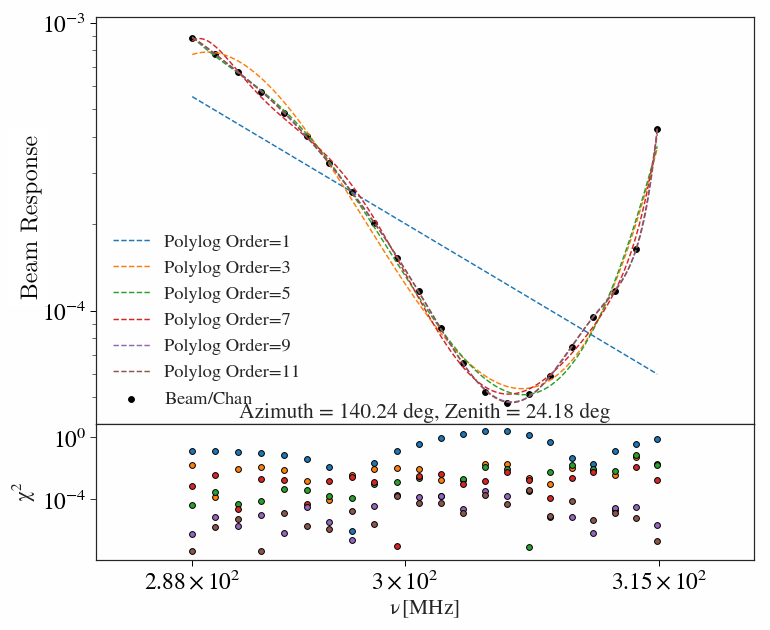}
            \caption{One of the more extreme examples of log-beam curvature across the 300\,MHz bandwidth. The individual black points are the coarse channels in the bandwidth. The beam response shows multiple changes in the gradient as well as a minima. Several log-space polynomials were fit to the log-beam coarse channels, in this figure we only show the odd ordered polynomials. These are represented by the coloured dashed lines. }
            \label{fig:kinky-oof}
\end{figure}

Equation \ref{eq:Sapp-mod} incorporates the intrinsic source SED model $S(\nu)$, and the spectral structure of the MWA tile beam response $B_{\theta,\phi}(\nu)$, which is determined at the source's position $(\theta,\phi)$. Additionally for snapshot observations we assume that at a fixed $(\theta,\phi)$, $B_{\theta,\phi}(\nu)$ is approximately constant over the duration of the observation. One type of $S_\textrm{app}$ model \textsc{calibrate} accepts is arbitrary order polylogarithmic functions. This is the model we use in this work:

\begin{equation}\label{eq:log-Sapp-mod}
    \log_{10}(S_{\textrm{app}}) = \sum^p_{i=0} a^\textrm{app}_i \left( \log_{10} \left(\frac{\nu}{\nu_0} \right) \right)^i
\end{equation}
$p$ is the polynomial order, and $\nu_0$ is the same reference frequency from Section \ref{Sec:PUMAcat-SEDmodels}. Equation \ref{eq:log-Sapp-mod} can also be expressed as a linear combination of two polylogarithmic functions $\log_{10}(S(\nu))$ and $\log_{10}(B_{\theta,\phi}(\nu))$ in log-space. $\log_{10}(B_{\theta,\phi}(\nu))$ is modelled as a polylogarithmic function to interpolate the MWA fine channel beam response for a particular source. This is done because the FEE tile beam model only models the coarse channels of the tile beam response \citep{BEAM2014,BEAM2016}. In many cases higher order polylogarithmic functions are required to accurately interpolate $\log_{10}(B_{\theta,\phi}(\nu))$ for a particular source. This is a result of some bright sources being located near MWA tile beam nulls where the beam response is changing quickly. An extreme example of a source near an MWA tile beam null is illustrated in Figure \ref{fig:kinky-oof}. 

The choice of polynomial order fit to $\log_{10}(B_{\theta,\phi}(\nu))$ depends on the location of the source in the beam response. In most cases the 11th order polynomial required to accurately model the beam response in Figure \ref{fig:kinky-oof} is not necessary. Generally a simple first order to fifth order polynomial is appropriate to model $\log_{10}(B_{\theta,\phi}(\nu))$. To choose the appropriate order fit, we calculate the $\chi^2$ value from Equation \ref{eq:chisqd}, with a $\sigma = 1$; we additionally determine the degrees of freedom $(\textrm{dof} = N - (p + 1))$. We use the minimisation of the $\chi^2$ and the $\textrm{dof}$ to select an the appropriate polynomial order fit to $\log_{10}(B_{\theta,\phi}(\nu))$. We limit the maximum order fit polynomial to $p \leq 11$, above this limit over-fitting starts to become an issue, and interpolation becomes increasingly inaccurate due to the Runge phenomenon \citep{Runge}.
 
\begin{figure*}[t]
    \begin{center}
        \begin{subfigure}[b]{0.495\textwidth}
            \centering
            \includegraphics[width=\textwidth]{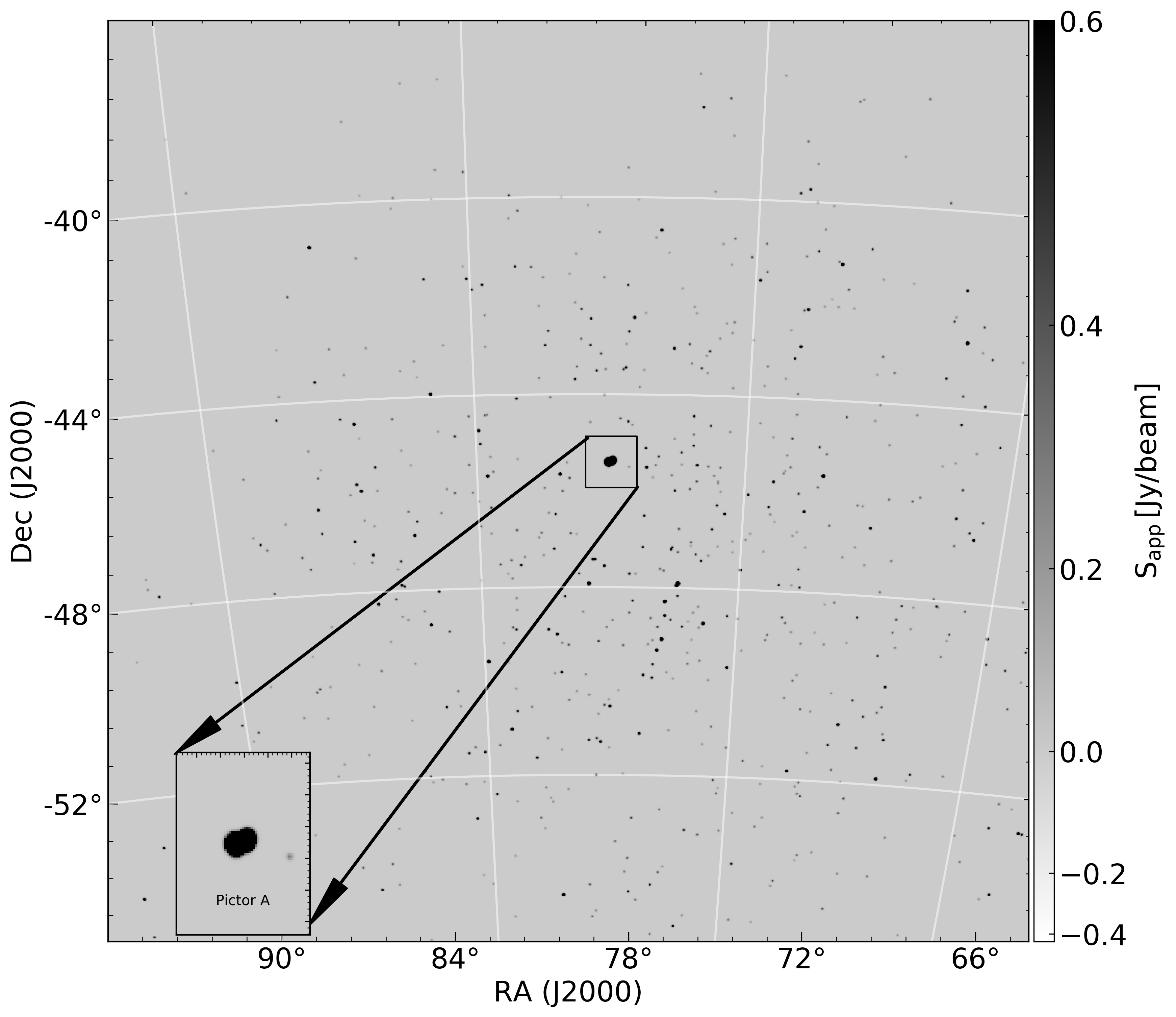}
            \caption[]%
            {{\small ObsA Main Lobe}}
            \label{fig:PicA-appsky-pb}
        \end{subfigure}
        \hfill
        \begin{subfigure}[b]{0.495\textwidth}
            \centering
            \includegraphics[width=\textwidth]{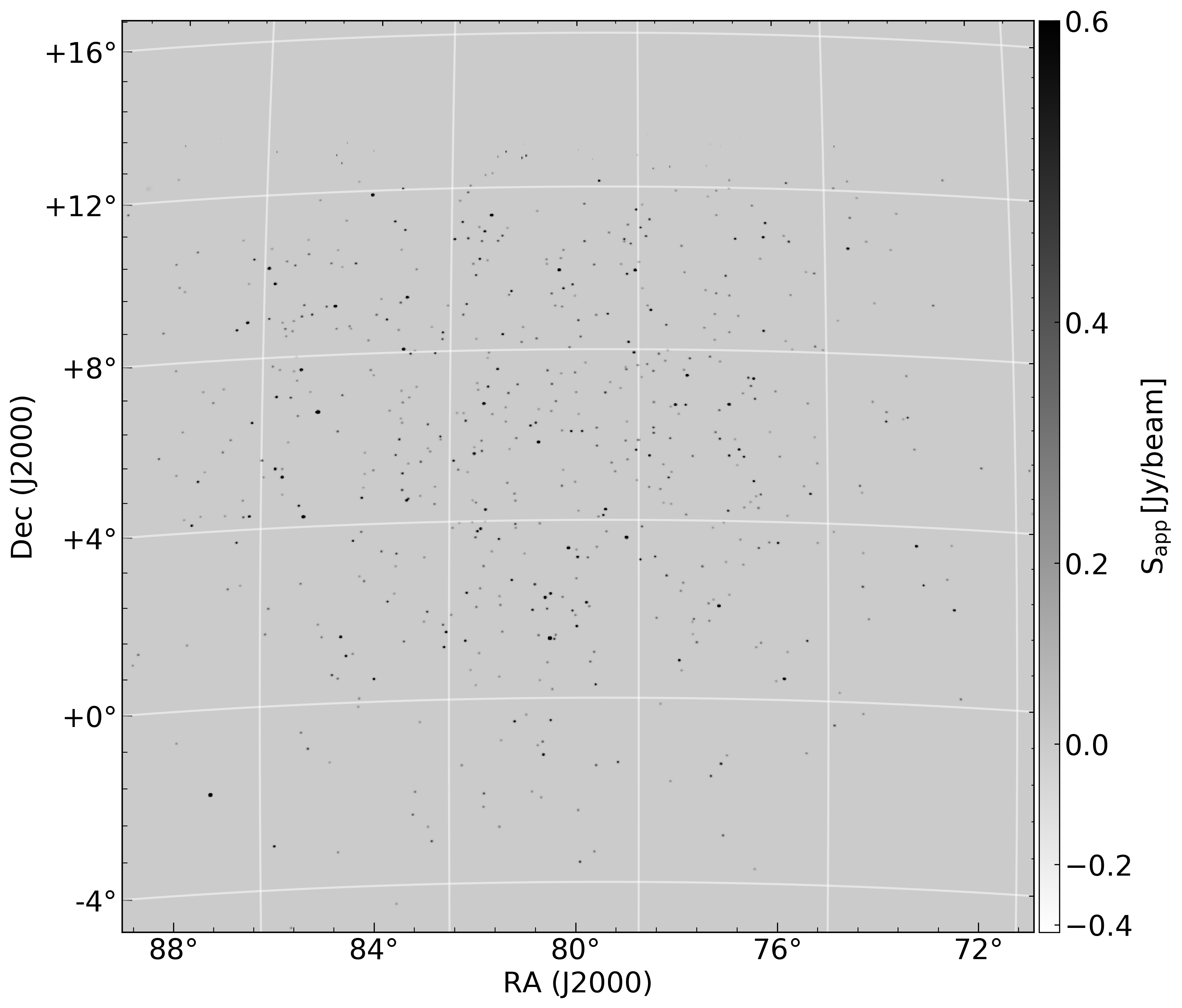}
            \caption[]%
            {{\small ObsA Grating Lobe}}
            \label{fig:PicA-appsky-GL}
        \end{subfigure}
        \caption[]
        {\small Image of the apparent sky-model for ObsA, Subfigure \ref{fig:PicA-appsky-pb} shows the main lobe of the observation, centred at $\textrm{RA}=79\degree.95$, $\textrm{DEC}=-45\degree.79$. PicA is visible in the enlarged box in the bottom left hand corner. Subfigure \ref{fig:PicA-appsky-GL} shows the prominent grating lobe for ObsA centred at $\textrm{RA}=79\degree.95$, $\textrm{DEC}=+5\degree$.} 
        \label{fig:PicA-appsky}
    \end{center}
\end{figure*}
 
Once the log-beam coarse channels have been fit for every source, we add the $\log_{10}(B_{\theta,\phi}(\nu))$ fit coefficients to the $\log_{10}(S(\nu))$ coefficients to determine the apparent coefficients in Equation \ref{eq:log-Sapp-mod}. The sources along with their fit coefficients are then written to a to a VOTable \citep{VO}. An example apparent sky-model at 300\,MHz can seen in Figure \ref{fig:PicA-appsky}, which shows the main lobe and the most prominent grating lobe from the apparent sky-model of the calibrator observation ObsA.

\subsection{Calibrating the Apparent Sky-Model}

The VOTable formatted apparent sky-model is converted into a text file, with a format specific to \textsc{calibrate}. This sky-model text document is a list from brightest to faintest of the 1500 selected sources. Each source in the list contains information about its RA and DEC, and the sources integrated apparent flux density at $300\,$MHz. If the source is a Gaussian it also provides the major and minor axes in arcseconds, as well as the position angle of the source relative to north in degrees. The apparent flux density coefficients for each source are additionally given.

This text file along with the observation measurement set are passed to \textsc{calibrate}, which is executed on uncalibrated visibilities and outputs the solutions to a binary file \citep{calibrate}. \textsc{applysolutions}\footnote{\textsc{calibrate} and \textsc{applysolutions} are apart of the same software package. They are both available at the same GitHub repository \href{https://github.com/ICRAR/mwa-reduce/commit/ef820c9ab62ad311dab17b0c3b6dce40002e4cfa}{mwa-reduce}. This repository is not publicly available, for access please contact the author. \url{https://github.com/ICRAR/mwa-reduce}} is used to apply calibration solutions to the same or other observations \citep{applysolutions}. After the initial calibration, we perform another round of RFI flagging. This flags calibration outliers and any additional RFI that is more prominent after initial calibration.

\subsection{All-Sky Imaging \& Self-Calibration}\label{Sec:allsky-image}

The apparent sky-model for calibrator observations provide a good first pass calibration of the observation visibilities, these can then be used to create deconvolved all-sky images \citep{CLEAN}. In this work all-sky imaging is performed with the software package \textsc{wsclean} \citep{WSCLEAN}. \textsc{wsclean} takes into account the w-terms for MWA observations due to the wide field of view, with a process called w-stacking \citep{w-stack}. The resulting CLEAN components generated from the all-sky images may then be used to perform another round of self-calibration.
 
To perform the all-sky imaging the visibilities are first phase shifted to zenith using the \textsc{wsclean} software \textsc{chgcentre} \citep{WSCLEAN}. The resulting zenith phase shifted visibilities reduce the w-terms, and are capable of producing an orthographic image of the entire sky. We then perform a shallow CLEAN with \textsc{wsclean}, with a threshold of $0.3$, an auto mask of $3.0$ and an mgain of $0.85$. Due to the high computation costs of performing deconvolution on high resolution all-sky images, we limit \textsc{wsclean} to $300,000$ minor iterations. The image size is additionally limited to $7000$ by $7000$ pixels due to memory constraints. 
 
\begin{figure*}[t]
    \begin{center}
        \begin{subfigure}[b]{0.495\textwidth}
            \centering
            \includegraphics[width=\textwidth]{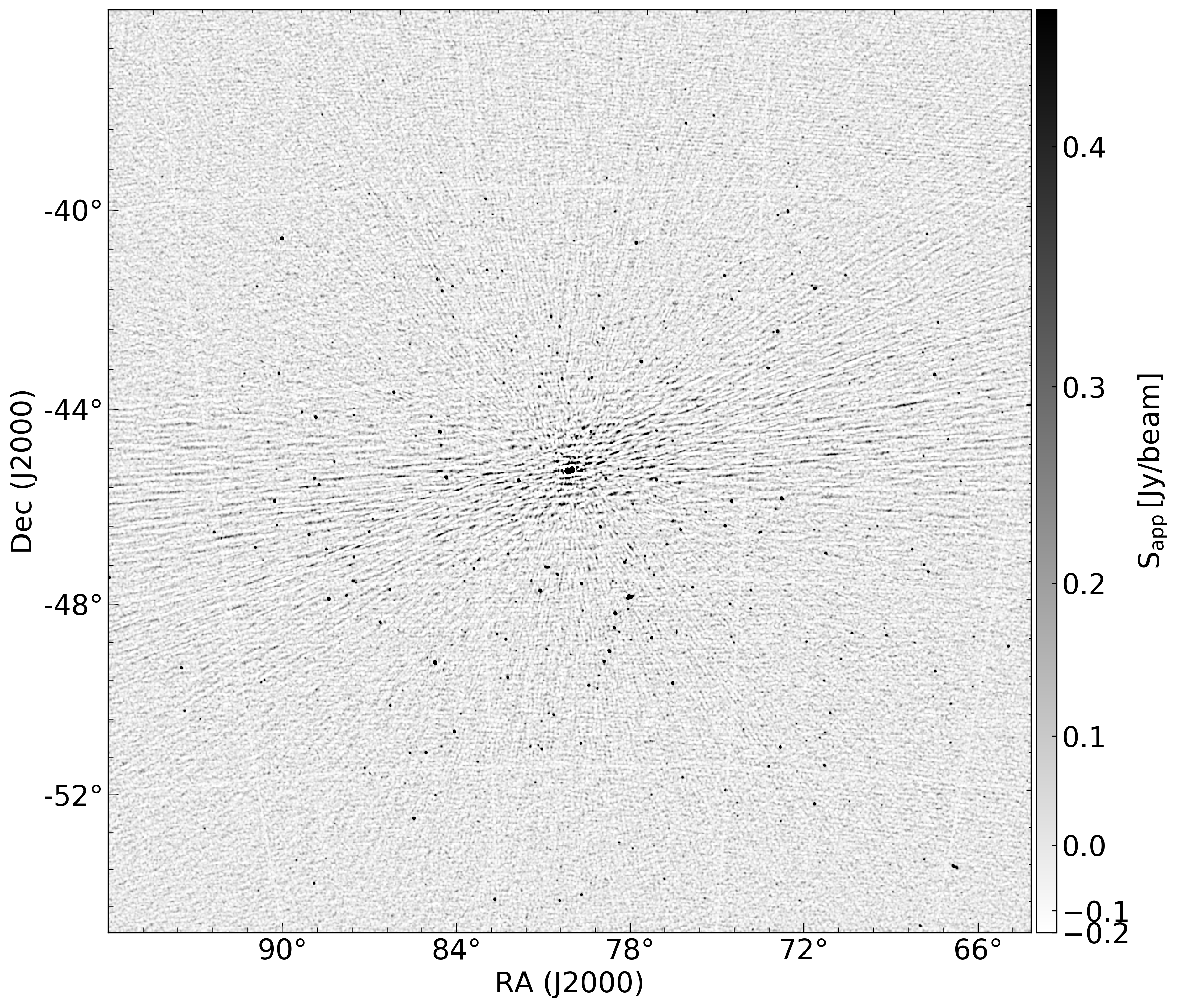}
            \caption[]%
            {{\small ObsA Main Lobe}}
            \label{fig:PicA-allsky-pb}
        \end{subfigure}
        \hfill
        \begin{subfigure}[b]{0.495\textwidth}
            \centering 
            \includegraphics[width=\textwidth]{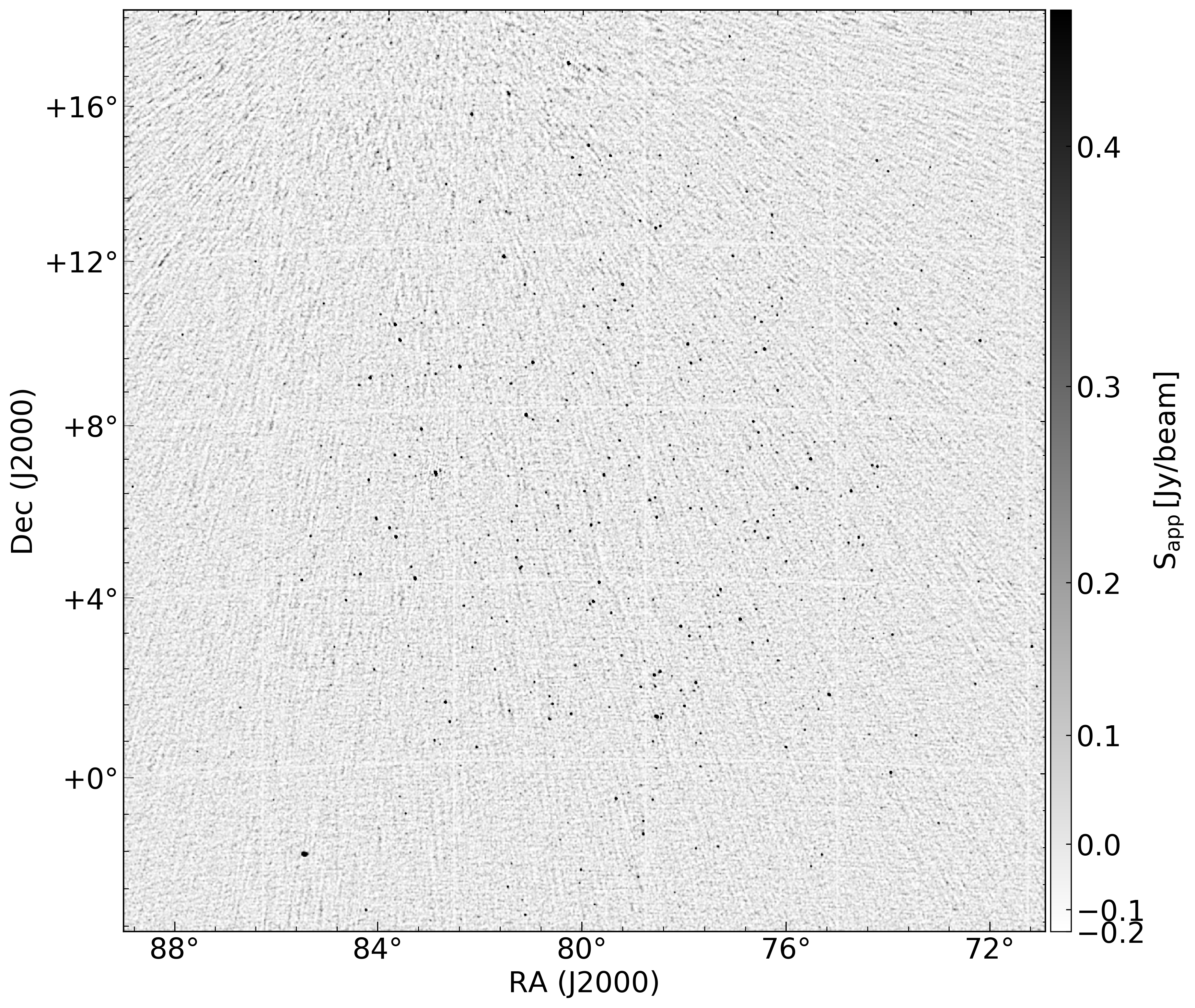}
            \caption[]%
            {{\small ObsA Grating Lobe}}
            \label{fig:PicA-allsky-GL}
        \end{subfigure}
        \caption[]
        {\small Apparent all-sky image of ObsA, presenting the main lobe, and the most prominent grating lobe. Subfigure \ref{fig:PicA-allsky-pb} shows the main lobe centred at $\textrm{RA}=79\degree.95$, $\textrm{DEC}=-45\degree.79$ with an rms of $86\,\textrm{mJy/beam}$. Subfigure \ref{fig:PicA-allsky-GL} shows the most prominent grating lobe centred at $\textrm{RA}=79\degree.95$, $\textrm{DEC}=+5\degree$ with an rms of $68\,\textrm{mJy/beam}$. There are additional grating lobes to the east and west of the main lobe which contain additional sources. Since the projection of this observation is significantly away from zenith, these grating lobes are significantly less prominent than the one shown in Subfigure \ref{fig:PicA-allsky-GL}. As such they were not included.}
        \label{fig:PicA-all-sky}
    \end{center}
\end{figure*}
 
Since the all-sky image is an orthographic projection, the projected diameter for the celestial sphere of unit radius is $114\degree.58$ (this is equivalent to two radians in degrees). Dividing this by the number of pixels along one dimension determines a pixel resolution of $\Delta\theta = 59\,\textrm{arcsec}$. Due to the constraints on the image size and thus the resolution, we use \textsc{wsclean} to convolve the PSF with a Gaussian of size $140\,\textrm{arcsec}$. This provides an effective PSF sampling rate of $\sim2.4$ pixels per PSF.

Examples of the main lobe, and the grating lobe from the apparent all-sky image of ObsA can be seen in Figure \ref{fig:PicA-all-sky}. Subfigure \ref{fig:PicA-allsky-pb} shows the main lobe which is dominated by the sidelobes of PicA. Subfigure \ref{fig:PicA-allsky-GL} is the most prominent grating lobe which contains numerous radio sources, some of which are very bright. The rms sensitivity of the main lobe and the grating lobe is $86\,\textrm{mJy/beam}$ and $68\,\textrm{mJy/beam}$ respectively. The sensitivity of the most prominent grating lobe is better than the main lobe due to the lack of high power PSF sidelobes from a bright source such as PicA.

During the CLEAN process \textsc{wsclean} generates a model where it stores the CLEAN component visibilities \citep{WSCLEAN}. \textsc{calibrate} can use the CLEAN component model visibilities to calibrate the data. For calibrator observations which have high signal to noise, the resulting gain and phase solutions are often better constrained. These solutions can be applied to the calibrator observation, or to a non-calibrator observation at the same pointing. After they have been applied an additional round of RFI/outlier flagging is performed.

\section{Main Lobe Imaging Strategy}\label{Sec:Imaging}

In radio interferometry, the main lobe of an observation (which determines the field of view), is typically the principal scientific region of interest. Since the main lobe is small relative to the entire sky, images of the main lobe can have higher resolution, and thus less restrictions to the deconvolution process discussed in Section \ref{Sec:allsky-image}. The presence of several highly sensitive grating lobes at 300\,MHz for MWA observations, means there are effectively several fields of view for a given observation. The radio sources which are present in these grating lobes produce sidelobe confusion that lowers the dynamic range of the main lobe. To properly image the main lobe, the contribution to the visibilities from the remaining parts of the sky need to be subtracted. In this section we describe the sky subtraction and imaging process for the main lobes of 300\,MHz observations. In principle this process can be generalised for any lobe, but for this work we only focus on the main lobe.

\begin{figure*}[t]
    \begin{center}
        \begin{subfigure}[b]{0.495\textwidth}
            \centering
            \includegraphics[width=\textwidth]{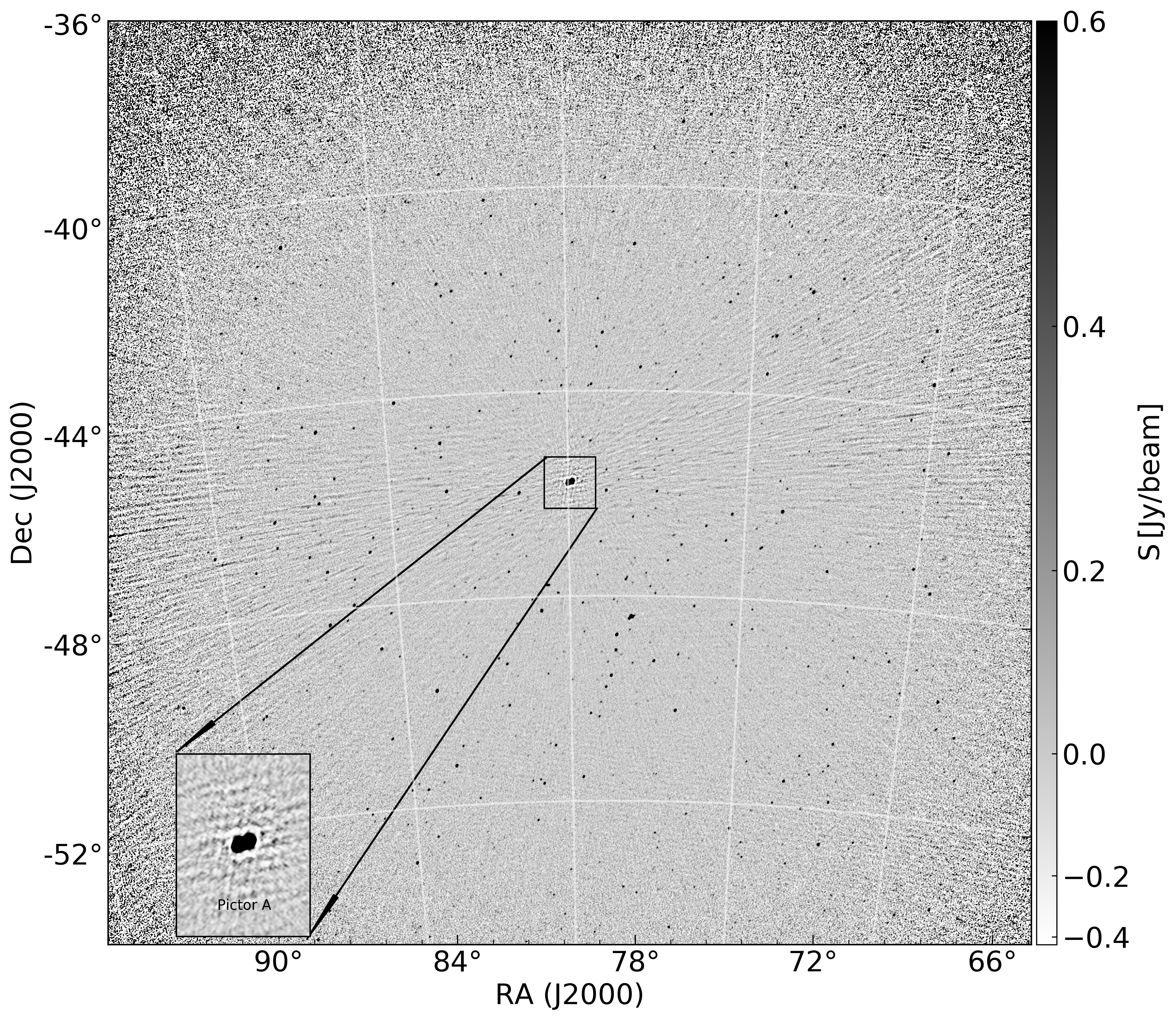}
            \caption[]%
            {{\small ObsA Main Lobe}}
            \label{fig:PicA-pb}
        \end{subfigure}
        \hfill
        \begin{subfigure}[b]{0.495\textwidth}
            \centering 
            \includegraphics[width=\textwidth]{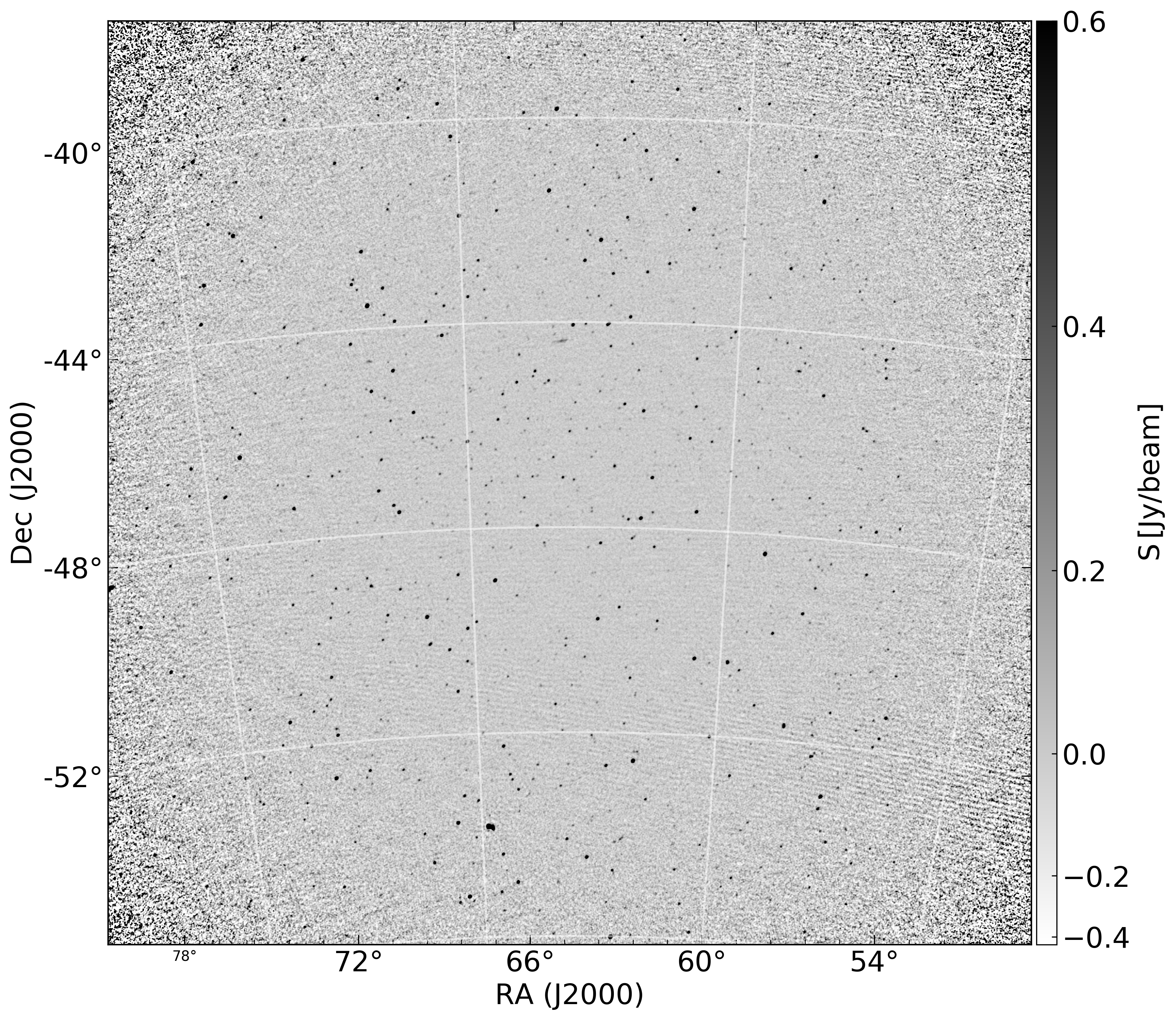}
            \caption[]%
            {{\small ObsB Main Lobe}}
            \label{fig:PicA-GL}
        \end{subfigure}
        \caption[]
        {\small Beam corrected Briggs $0.0$ weighted main lobe images for ObsA and ObsB in Subfigure (a) and (b) respectively. In Subfigure (a) the enlarged region shows PicA which at the resolution of this image is unresolved. Faint sidelobe artefact can be seen in both images, where the rms for Subfigure (a) is $56\,\textrm{mJy/beam}$ and $31\,\textrm{mJy/beam}$ for Subfigure (b). The pixel scale for both images is $18\,\textrm{arcsec}$. The deeper rms for Subfigure (b) is a result of the absence of PicA in the main lobe.}
        \label{fig:ObsA-ObsB-PB}
    \end{center}
\end{figure*}

\subsection{Sky Subtraction Algorithm}\label{Sec:sky-sub}

To image the main lobe we developed a method which uses \textsc{wsclean} to remove the sky contribution from the visibilities. In the process of imaging the entire sky (described in Section \ref{Sec:allsky-image}), \textsc{wsclean} outputs the CLEAN model components to an image with the same dimensions as the output all-sky image \citep{WSCLEAN}. Using the \textsc{wsclean} \textsc{predict} function the visibilities of this image can be estimated, and written to the model column of the observation measurement set. To separate the main lobe contribution to the visibilities from the rest of the sky, we mask the main lobe in the model image by setting all pixel values to zero. We then run \textsc{predict} function on the masked image. Using the \textsc{wsclean} \textsc{taql}\footnote{This comes with \textsc{wsclean} and is an SQL based database command.} command the model visibilities are subtracted from the calibrated visibilities. We then perform another round of all-sky imaging, this will CLEAN the sources that were missed in the original run (at the cost of extra computation). This process can be iteratively repeated, but the default number of iterations is one. After each imaging stage we flag any additional RFI and calibration outliers. 

\subsection{Main Lobe Image Parameters}\label{Sec:Main-lobe}

After the sky-subtraction process is performed the main lobe of the observation is generated using \textsc{wsclean}. Main lobe images are by default generated with a Briggs weighting of $0.0$, using the \textsc{wslean} multi-scale CLEAN option \citep{WSC-multi}. The additional default imaging options are an auto threshold of $1$, with an auto mask of $3$, and an mgain of $0.5$ for $500,000$ minor iterations. Output images have dimensions of $5000$ by $5000$ pixels, with a pixel scale of $\Delta\theta \sim 18\,\textrm{arcsec}$.

\section{Results}\label{Sec:Results}

In this section we present the output main lobes images for the calibrator observation ObsA, and a non-calibrator observation ObsB, which we calibrated using solutions derived from ObsA. We apply the process outlined in Sections \ref{Sec:Calibration} and \ref{Sec:Imaging} to these two example observations to illustrate the method, the main lobe images of ObsA and ObsB are shown in Figure \ref{fig:PicA-pb} and Figure \ref{fig:PicA-GL} respectively.

\begin{figure*}[t]
    \begin{center}
        \begin{subfigure}[b]{0.495\textwidth}
            \centering 
            \includegraphics[width=\textwidth]{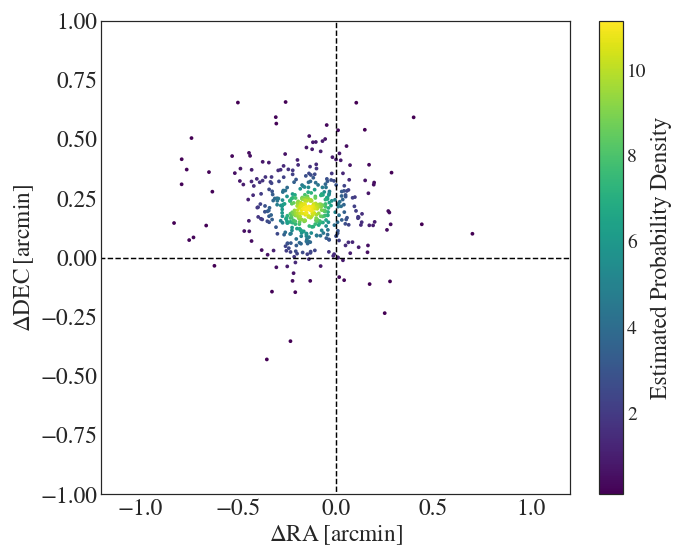}
            \caption[]%
            {{\small ObsA}}
            \label{fig:PicA-astro}
        \end{subfigure}
        \begin{subfigure}[b]{0.495\textwidth}
            \centering 
            \includegraphics[width=\textwidth]{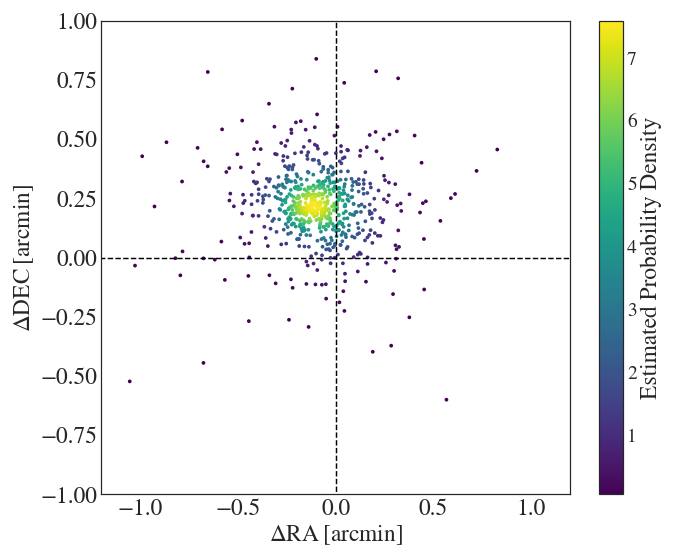}
            \caption[]%
            {{\small ObsB}}
            \label{fig:ObsB-astro}
        \end{subfigure}
        \caption[]
        {\small Difference in the RA and DEC between the model and the measured sources for ObsA (Subfigure \ref{fig:PicA-astro}) and ObsB (Subfigure \ref{fig:ObsB-astro}). The dashed black lines for both figures show how far the sources deviate from an offset of zero. The colour bar shows the estimate probability density for both figures.}
        \label{fig:Astro}
    \end{center}
\end{figure*}

\begin{figure*}[t!]
    \begin{center}
        \begin{subfigure}[b]{0.495\textwidth}
            \centering 
            \includegraphics[width=\textwidth]{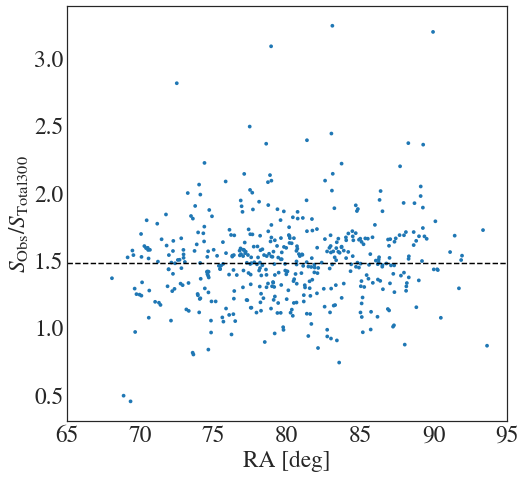}
            \caption[]%
            {{\small ObsA $S_\textrm{Obs}/S_\textrm{Total300}$ Vs RA}}
            \label{fig:ObsA-RA}
        \end{subfigure}
        \begin{subfigure}[b]{0.495\textwidth}
            \centering 
            \includegraphics[width=\textwidth]{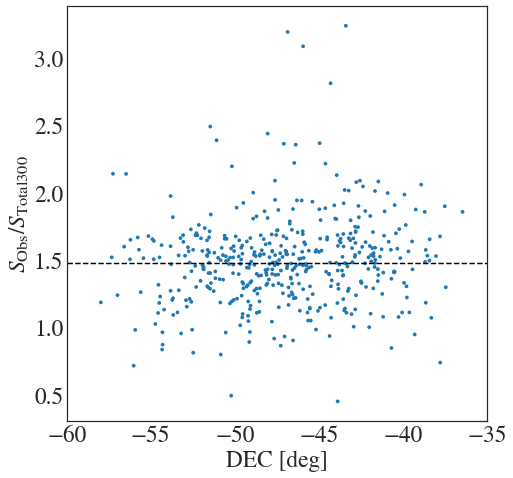}
            \caption[]%
            {{\small ObsA $S_\textrm{Obs}/S_\textrm{Total300}$ Vs DEC}}
            \label{fig:ObsA-DEC}
        \end{subfigure}
        \begin{subfigure}[b]{0.495\textwidth}
            \centering 
            \includegraphics[width=\textwidth]{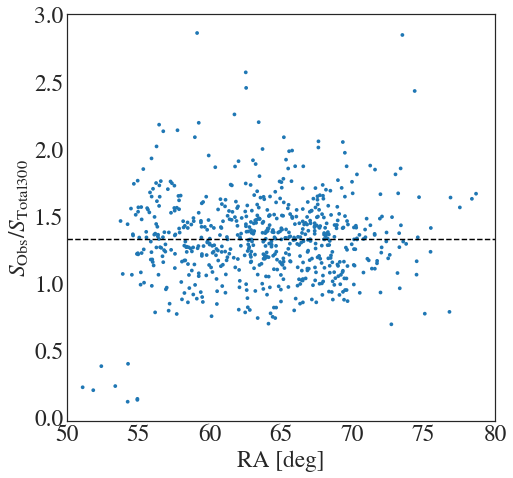}
            \caption[]%
            {{\small ObsB $S_\textrm{Obs}/S_\textrm{Total300}$ Vs RA}}
            \label{fig:ObsB-RA}
        \end{subfigure}
        \begin{subfigure}[b]{0.495\textwidth}
            \centering 
            \includegraphics[width=\textwidth]{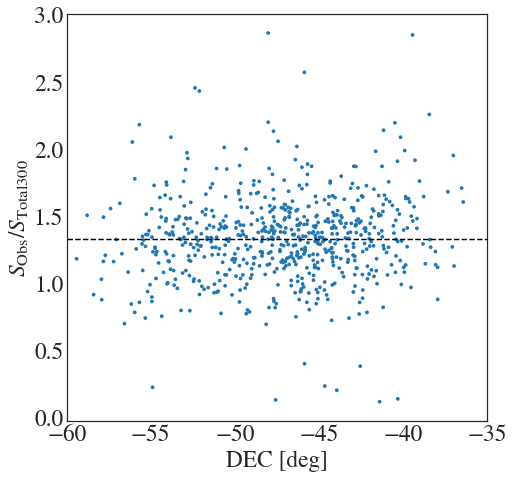}
            \caption[]%
            {{\small ObsB $S_\textrm{Obs}/S_\textrm{Total300}$ Vs DEC}}
            \label{fig:ObsB-DEC}
        \end{subfigure}
        \caption[]
        {\small Scatter plot of the flux density ratio for ObsA and ObsB against RA (Subfigures \ref{fig:ObsA-RA} and \ref{fig:ObsB-RA}) and DEC (Subfigures \ref{fig:ObsA-DEC} and \ref{fig:ObsB-DEC}). The dashed black line indicates the median flux density ratio for both ObsA and ObsB.}
        \label{fig:flux-scale}
    \end{center}
\end{figure*}

\subsection{ObsA \& ObsB Main Lobe Images}\label{Sec:ObsA-ObsB-images}

Comparing the sensitivity of the ObsA main lobe image from Figure \ref{fig:PicA-pb} to Figure \ref{fig:PicA-allsky-pb}, the former clearly has many more visible point sources. The rms in Figure \ref{fig:PicA-pb} is $56\,\textrm{mJy/beam}$ compared to $86\,\textrm{mJy/beam}$ in Figure \ref{fig:PicA-allsky-pb}. The majority of the improvement in sensitivity comes from eliminating sidelobe confusion, through the application of the sky-subtraction method, and by performing a deeper CLEAN on PicA. Additional gains in sensitivity come from using a Brigg's weighting of $0.0$, which balances resolution for an increase in sensitivity. Though it should be noted, that due to the Brigg's weighting of $0.0$ the size of the major axis for the restoring beam in Figure \ref{fig:PicA-pb} is $\sim2.4\,\textrm{arcmin}$. This is approximately the same size as the restoring beam in Figure \ref{fig:PicA-allsky-pb}. Therefore for this comparison, this might have less of a contribution to the relative increase in sensitivity, than the removal of the sidelobe confusion. Further gains in sensitivity come from removing RFI.

For ObsA the total flagged visibilities percentage is $\sim46\%$. Referring to Equations \ref{eq:sensitivity} and \ref{eq:sensitivity-flgd} of Section \ref{sec:sensitivity} in the appendix, for a naturally weighted observation the best sensitivity for a snapshot 300\,MHz observation is $\sim26\,\textrm{mJy/beam}$. In Subfigure \ref{fig:PicA-pb} there are still sidelobes present for PicA, which will be contributing to the noise through sidelobe confusion. For ObsB since PicA is not present in the main lobe, the main lobe sensitivity is $31\,\textrm{mJy/beam}$. For the non-calibrator observation ObsB, when accounting for the different weighting schemes and potential flux scale calibration errors, this sensitivity limit is close to the theoretical prediction.

\subsection{ObsA \& ObsB Astrometry}

With the deep main lobe images for ObsA and ObsB we investigate the accuracy of the source positions relative to the Total300 catalogue. Using the source finder \textsc{aegean} we create source lists for ObsA and ObsB \citep{Aegean,Aegean2}. The total number of sources found by \textsc{aegean} were $457$ for ObsA and $656$ for ObsB respectively. These source lists are then cross-matched with the Total300 catalogue at a angular separation of two arcminutes. The cross-matching is performed by the astronomy software \textsc{topcat}\footnote{\url{http://www.star.bris.ac.uk/~mbt/topcat/}} \citep{Topcat}. The resulting cross-matched catalogues for ObsA and ObsB constitute a completeness of $\sim99\%$. Due to the low sensitivity threshold for the ObsA and ObsB main lobe images, we should expect to see $100\%$ cross-match with Total300 and the \textsc{aegean} source catalogues. Sources that do not have a match are potentially artefacts, or have peaks in their spectrum at higher frequencies. High frequency peaked sources, would be fainter at lower frequencies.
 
Using the cross-matched catalogue we determine the angular offsets for the sources in RA $(\Delta\textrm{RA})$ and DEC $(\Delta\textrm{DEC})$. The resulting astrometry plots for ObsA and ObsB are given in Figure \ref{fig:Astro}. There is a clear bulk offset of $\sim 12\,\textrm{arcsec}$ for both ObsA and ObsB in Figure \ref{fig:Astro}. This offset is smaller than the pixel scale, and is close to the expected $10\,\textrm{arcsec}$ ionospheric offset at $300\,$MHz \citep{ion-offset}. The standard deviation for $\Delta\textrm{RA}$ for ObsA and ObsB is $7.8\,\textrm{arcsec}$ and $10.2\,\textrm{arcsec}$ respectively. The standard deviation for $\Delta\textrm{DEC}$ for ObsA and ObsB is $5.4\,\textrm{arcsec}$ and $7.2\,\textrm{arcsec}$ respectively. There does not appear to be any correlation between the RA and DEC offsets.

\subsection{ObsA \& ObsB Flux Scale}

Using the cross-matched catalogues, the flux scale for ObsA and ObsB was determined by taking the ratio of the measured integrated flux density $(S_\textrm{Obs})$ determined by \textsc{aegean}, to the Total300 model integrated flux density $(S_\textrm{Total300})$. The median flux density ratio for ObsA is $1.48\pm0.26$, and $1.33\pm0.25$ for ObsB respectively. The deviation from a unitary flux density ratio indicates there is an error in the flux scale calibration. This could potentially result from underestimating the total flux density in the calibrator observation ObsA. The scatter plots in Figure \ref{fig:flux-scale} show the flux density ratio plotted against RA and DEC for both ObsA and ObsB. There does not appear to be any systematic bias in the flux density in relation to either RA or DEC for either observation. In Figure \ref{fig:ObsB-RA} and \ref{fig:ObsB-DEC} there appears to be a cluster of outlier sources with flux density ratios less than $0.5$. These sources are close to the edge of the main lobe, which could be indicative of an incorrect primary beam correction.

\section{Discussion \& Conclusion}\label{Sec:Discussion-Conlcusion}

This work demonstrates a successful calibration and imaging strategy for 300\,MHz MWA observations, and the processing pipeline used in this work is publicly available in the GitHub repository \href{https://github.com/JaidenCook/300-MHz-Pipeline-Scripts}{\textsc{S300-pipeline}}\footnote{https://github.com/JaidenCook/300-MHz-Pipeline-Scripts}. To date this pipeline has successfully processed over 15 300\,MHz observations, and is flexible enough to also process MWA observations at lower frequencies. The method outlined in this paper, works best when calibrating a bright calibrator source, and then transferring the solutions to another observation at the same pointing. Some observations of relatively low brightness fields can be calibrated without a calibrator observation, but in practice this is very difficult.

There are many difficulties associated with processing 300\,MHz MWA data. In particular RFI was a larger issue than anticipated. Each processing step required additional flagging to remove the RFI contribution to the visibilities. Additionally some observations show evidence of intervening satellites which either reflect or transmit at the lower frequency coarse channels in the 300\,MHz band. Issues with the flux scale calibration can arise when calibrating on certain calibrator sources such as Hydra A. 

The biggest improvement that can be made to this work is the sky-model. With the later release of GLEAM data, we can use PUMA to cross match the GP, SMC, and LMC to obtain higher frequency information for these regions. This will allow for interpolation of the 300\,MHz flux, extrapolation can be unreliable for source flux density estimation as can be seen from some sources in GLEAM\_Sup. Additionally data releases of high frequency surveys such as the Rapid ASKAP Continuum Survey \citep[RACS;][]{RACS} which cover the same area of the sky as GLEAM, will help to fill in the SEDs for sources in the sky-model \citep{RACS}. The RACS first data release is for band one which is centred at a frequency of $887.5\,$MHz with a bandwidth of $288\,$MHz. This will help to constrain the higher frequency end of the sky-model. With a more comprehensive sampling of the $72-1400\,$MHz frequency range we can refit the SED models for the Total300 catalogue. Additional improvements to the low-frequency end will come with the release of the GLEAM-X data, which will offer better resolutions. Making a more reliable processing pipeline for extended MWA observations at 300\,MHz.

\begin{acknowledgements}
I would like to thank C. M. Trott who both gave me advice in relation to the work in this paper.

This work was supported by resources provided by the Pawsey Supercomputing Centre with funding from the Australian Government and the Government of Western Australia. This scientific work makes use of the Murchison Radio-astronomy Observatory, operated by CSIRO. We acknowledge the Wajarri Yamatji people as the traditional owners of the Observatory site. Support for the operation of the MWA is provided by the Australian Government (NCRIS), under a contract to Curtin University administered by Astronomy Australia Limited. We acknowledge the Pawsey Supercomputing Centre which is supported by the Western Australian and Australian Governments.

\end{acknowledgements}

\begin{appendix}

\section{Theoretical 300\,MHz Sensitivity Limit}\label{sec:sensitivity}

The radiometer equation can be used to estimate the sensitivity limit for an MWA snapshot observations at $300\,\rm{MHz}$:

\begin{align}\label{eq:sensitivity}
    \sigma_{S_{300}} = \frac{2k_bT}{A_{\rm{eff}}}\sqrt{\frac{1}{N(N-1) \Delta\nu \tau_0}}
\end{align}

Where $\tau_0=120$ is the snapshot observation time in seconds, $T_{\textrm{sys}} = T_{\textrm{sky}} + T_{Rc}$ the sky temperature at $300\,\textrm{MHz}$ summed with the receiver temperature, $A_{\textrm{eff}}=4.75\,\textrm{m}^2$ is the effective tile area \citep{Eff-area}, $N=128$ is the number of tiles, $k_b$ is Boltzmann's constant, and $\Delta\nu = 30.72\,\textrm{MHz}$ is the bandwidth. The sky temperature is given by $T_{\textrm{sky}} = 60\lambda^{2.25}\,\textrm{K}$ \citep{MWA2}, hence at $300\,\textrm{MHz}$, $T_{\textrm{sky}} = 60\,\textrm{K}$. Finally $T_{Rc} = 180\,\textrm{K}$, therefore $T_{\textrm{sys}} = 240\,\textrm{K}$. Using these values the sensitivity is estimated to be $\sigma_{S_{300}} \approx 19\,\textrm{mJy}$. This is the best case scenario assuming that no fine channels are flagged. 
  
Equation \ref{eq:sensitivity} is the estimated RMS for a naturally weighted set of visibilities. In reality there are many kinds of weighting schemes that can be applied to the data, these affect the sensitivity. An in depth derivation of interferometric sensitivity and how weighting schemes can affect the RMS can be found in Section 6.2.3 of \citep{Radio-synthesis}.

\begin{equation}\label{eq:sensitivity-gen}
    \sigma_{S_{300}} = \frac{2k_bT}{A_{\textrm{eff}}} \sqrt{\frac{1}{N(N-1)\Delta\nu \tau_0}} \frac{w_{\textrm{mean}}}{w_\textrm{rms}}
\end{equation}

Equation \ref{eq:sensitivity-gen} represents the general form which accounts for the weighting of the data. The general from is Equation \ref{eq:sensitivity} scaled by the ratio of the mean weighting to the RMS of the weightings.

\subsection{Sensitivity of Flagged Data}\label{sec:sensitivity-flagged-data}

Following a similar argument we can likewise express $\sigma^\prime_{S_{300}}$ as:
\begin{equation}
    \sigma^\prime_{S_{300}} = \frac{2k_bT}{A_{\rm{eff}}}\sqrt{\frac{1}{n_d(1-R_f) \tau_\alpha \Delta\nu_\alpha}}
\end{equation}
We then have the relationship:
\begin{equation}\label{eq:sensitivity-flgd}
    \sigma^\prime_{S_{300}} = \sigma_{S_{300}}\frac{1}{\sqrt{1-R_f}}
\end{equation}

Using Equation \ref{eq:sensitivity-flgd} if we know the percentage of data flagged we can estimate the expected theoretical sensitivity of the observation. This does not take into consideration the different weighting schemes that are applied to the independent visibility data points.

\section{Polylogarithmic Coordination Transformation Proof}\label{sec:poly-log-proof}

\begin{align}
    f(\nu) &= \sum^p_{i=0} a_i \left( \log\left( \frac{\nu}{\nu_a}\right) \right)^i \\
    g(\nu) &= \sum^p_{i=0} b_i \left( \log\left( \frac{\nu}{\nu_b}\right) \right)^i
\end{align}

In the above equations $\nu_a$ and $\nu_b$ are the normalisation constants for their respective polylogarithmic functions. Now consider the scenario where $f(\nu)=g(\nu)\: \forall \: \nu \: \in \: \mathbb{R}$, but $\nu_a \neq \nu_b$ and $a_i \neq b_i \: \forall \: i$. 

\paragraph{Proposition:} There should exist a transform of the coefficients of $\mathbf{a}$ from the space $\nu/\nu_a$ to the space $\nu/\nu_b$

There should exist an expression for each coefficient $b_i$ as a linear combination of the product of the coefficients $a_i$, $\log(\nu_b/\nu_a)$, and the binomial coefficients:

\begin{equation}
    b_l = \sum^p_{i=l} a_i {{i}\choose{i-l}} \left(\log\left( \frac{\nu_b}{\nu_a} \right)\right)^{i-l}
\end{equation}

\paragraph{Proof:} Here we will show through induction how to express equation 7 as a linear combination of the terms $\log(\nu/\nu_b)$ and hence derive an expression for each of the coefficients $b_i$. First we let $\log(\nu/\nu_a) = \log(\nu/\nu_b) + \log(\nu_b/\nu_a) = x + y$, we can then rewrite equation 7:

\begin{multline}
    f(x(\nu)) = \sum^p_{i=0}a_i\left(x + y \right)^i = a_0 + a_1\left(x + y \right) + \cdots + \\ a_{p-1}\left(x+y\right)^{p-1} + a_p \left(x+y\right)^p
\end{multline}

We can expand each term in the sum $\left(x + y\right)^i$ through a binomial expansion, and hence rewrite each $\left(x + y \right)^i$ term as a sum:

\begin{equation}
\begin{split}
f(x(\nu)) &= \sum^p_{i=0} a_i  \left[{{i}\choose{0}} x^i + {{i}\choose{1}} x^{i-1} y + \cdots \right. \\
    &\left.+ {{i}\choose{i-1}}xy^{i-1} + {{i}\choose{i}} y^i  \right] \\
    &= \sum^p_{i=0} a_i \sum^i_{j=0} {{i}\choose{j}} x^{i-j}y^j
\end{split}
\end{equation}

By factoring out the zeroth $x$ terms we can rewrite the expression in equation 12:

\begin{multline}
    f(x(\nu)) =  \sum^p_{i=0} a_i \left[\sum^{i-1}_{j=0} {{i}\choose{j}} x^{i-j}y^j + {{i}\choose{i}} y^i \right]\\
    =\sum^p_{i=0} a_i {{i}\choose{i}} y^i + \sum^p_{i=1} a_i \sum^{i-1}_{j=0} {{i}\choose{j}} x^{i-j}y^j
\end{multline}

Since all the zeroth order $x$ terms have been factored out, the new inner sum reduces by $1$, and the outer sum subsequently increments by $1$ since the inner sum cannot start at $-1$. This factorisation process can be extended to each successive lowest order $x$ term, to generally prove this, consider the arbitrary step $k$ which is defined below:

\begin{equation}
\begin{split}
    &\sum^p_{i=k} a_i \sum^{i-k}_{j=0} {{i}\choose{j}} x^{i-j}y^j =\\
    &\sum^p_{i=k} a_i \left[ {{i}\choose{0}} x^i + {{i}\choose{1}} x^{i-1} y + \cdots  + {{i}\choose{i-k}} x^ky^{i-k} \right]
\end{split}
\end{equation}

We see in equation 15 that similar to the form written in equation 11, that if we let $k=0$ we reduce to the entire sum. Now we factor out the kth order $x$ term from equation 15:

\begin{align}
    &\sum^p_{i=k} a_i \sum^{i-k}_{j=0} {{i}\choose{j}} x^{i-j}y^j = \\
    &\sum^p_{i=k} a_i \left[\sum^{i-k-1}_{j=0} {{i}\choose{j}} x^{i-j}y^j + {{i}\choose{i-k}} x^ky^{i-k} \right] =\\
    &\left( \sum^p_{i=k} a_i {{i}\choose{i-k}} y^{i-k} \right)x^k + \sum^p_{i=k+1} a_i \sum^{i-k-1}_{j=0} {{i}\choose{j}} x^{i-j}y^j
\end{align}

Again we see that this factorisation reflects that of the zeroth order term. If we let $p=k+1$, hence $k=p-1$, then we retrieve the last two terms of the factorisation, for the highest and second highest orders of $x$:

\begin{multline}
    \sum^p_{i=k} a_i \sum^{i-k}_{j=0} {{i}\choose{j}} x^{i-j}y^j = \\ \left( \sum^p_{i=p-1} a_i {{i}\choose{i-(p-1)}} y^{i-(p-1)} \right)x^{p-1} + a_p {{p}\choose{0}} x^p
\end{multline}

We are now in a position to express equation 10 as a linear combination of $x$:

\begin{align}
    &f(x(\nu)) = \sum^p_{i=0} a_i {{i}\choose{i}} y^i + \cdots \\ 
    &+ \left( \sum^p_{i=l} a_i {{i}\choose{i-l}} y^{i-l} \right) x^l + \cdots + a_p {{p}\choose{0}} x^p \\
    &= b_0 + \cdots + b_l x^l + \cdots + b_p x^p = g(x(\nu))
\end{align}

It is clear to see how the coefficients of $f(x(\nu))$ map to the coefficients of $g(x(\nu))$, specifically the total sum can be expressed as:

\begin{align}
    g(x(\nu)) &= \sum^p_{l=0}\left(\sum^n_{i=l} a_i {{i}\choose{i-l}} y^{i-l} \right) x^l \\
    &= \sum^p_{l=0}\left(\sum^p_{i=l} a_i {{i}\choose{i-l}} \log^{i-l}\left(\frac{\nu_b}{\nu_a}\right) \right) \log^l\left(\frac{\nu}{\nu_b}\right)
\end{align}

And hence for an arbitrary coefficient $b_l$ we can express the transformation as:

\begin{equation}
    \therefore b_l = \sum^p_{i=l} a_i {{i}\choose{i-l}} \log^{i-l}\left(\frac{\nu_b}{\nu_a}\right)
\end{equation}

\subsection{Matrix Representation of the Polylogarithmic Coefficient Transformation Function}

We can see the emergent pattern where the sum of binomial coefficients in equation 23 is the linear combination of the right diagonals on pascals triangle, where $l$ is the order.

\begin{center}
\Longstack[l]{
p=0\\
p=1\\
p=2\\
p=3\\
p=4\qquad\ \\
}
\Longstack{
1\\
1\x 1\\
1\x 2\x 1\\
1\x 3\x 3\x 1\\
1\x 4\x 6\x 4\x 1\\
\overline{0\x 1\x 2\x 3\x 4}
}
\end{center}

We can use this pattern to define a transformation matrix that will operate on the vector of coefficients:

\begin{equation*}
\mathbf{a} = 
\begin{pmatrix}
a_0 & a_1 & \cdots & a_p
\end{pmatrix}
\end{equation*}
 
Thus transforming vector $\mathbf{a}$ into vector $\mathbf{b}$:

\begin{equation*}
\mathbf{b} = 
\begin{pmatrix}
b_0 & b_1 & \cdots & b_p
\end{pmatrix}
\end{equation*}

Consider the upper triangular matrix $\mathbf{P}_{n,n}$ where the dimensions of the triangular matrix $n$ are defined as the order $p+1$, this is equal to the number of coefficients in an arbitrary polynomial.

\begin{equation*}
\mathbf{P}_{n,n} = 
\begin{pmatrix}
{0}\choose{0} & {1}\choose{0} & {2}\choose{0} & \cdots & {p}\choose{0} \\
 & {1}\choose{1} & {2}\choose{1} & \cdots & {p}\choose{1} \\
  &  & \ddots & \ddots & \vdots  \\
  &  &  & \ddots & {p}\choose{p-1}  \\
0 &  &  &  & {p}\choose{p} 
\end{pmatrix}
\end{equation*}

In this matrix row consists of the diagonal elements of pascals triangle filling in from right to left, with nill entries being represented by zeros. The next important matrix is the polynomial matrix, which is also an upper triangular matrix, this is represented below:

\begin{equation*}
\mathbf{T}_{n,n} = 
\begin{pmatrix}
1 & y & y^2 & \cdots & y^p \\
 & 1 & y & \cdots & y^{p-1} \\
  &  & \ddots & \ddots & \vdots  \\
  &  &  & \ddots & y  \\
0 &  &  &  & 1 
\end{pmatrix}
\end{equation*}

In the matrix $\mathbf{T}_{n,n}$, the variable $y$ takes on the same value as it did in the previous section. The Hadmard product (element by element product) of the matrices $\mathbf{P}$ and $\mathbf{T}$, produces the polylogarithmic coefficient transformation matrix:

\begin{equation*}
\mathbf{A}_{n,n} = 
\begin{pmatrix}
{0}\choose{0} & {{1}\choose{0}}y & {{2}\choose{0}}y^2 & \cdots & {{p}\choose{0}}y^p \\
 & {1}\choose{1} & {{2}\choose{1}}y & \cdots & {{p}\choose{1}}y^{p-1} \\
  &  & \ddots & \ddots & \vdots  \\
  &  &  & \ddots & {{p}\choose{p-1}}y  \\
0 &  &  &  & {p}\choose{p} 
\end{pmatrix}
\end{equation*}

Thus the transformation can be represented by:

\begin{align}
    \left(\mathbf{P} \circ \mathbf{T}\right)\mathbf{a}^\top &= \mathbf{b}^\top \\
    \therefore \mathbf{A}\mathbf{a}^\top &= \mathbf{b}^\top
\end{align}

Insert example with second order polylogarithmic functions.

\paragraph{Example (Second order Polylogarithmic Functions):}

\begin{align}
    f(\nu) &= a_0 + a_1 \left( \log\left( \frac{\nu}{\nu_a}\right) \right) + a_2 \left( \log\left( \frac{\nu}{\nu_a}\right) \right)^2\\
    g(\nu) &= b_0 + b_1 \left( \log\left( \frac{\nu}{\nu_b}\right) \right) + b_2 \left( \log\left( \frac{\nu}{\nu_b}\right) \right)^2
\end{align}

Hence:

\begin{equation*}
\mathbf{P}_{n,n} = 
\begin{pmatrix}
  1 & 1 & 1\\
  0 & 1 & 2\\
  0 & 0 & 1
\end{pmatrix}
\end{equation*}

\begin{equation*}
\mathbf{T}_{n,n} = 
\begin{pmatrix}
  1 & \left( \log\left( \frac{\nu}{\nu_a}\right) \right) & \left( \log\left( \frac{\nu}{\nu_a}\right) \right)^2\\
  0 & 1 & \left( \log\left( \frac{\nu}{\nu_a}\right) \right)\\
  0 & 0 & 1
\end{pmatrix}
\end{equation*}

Thus the polylogarithmic coefficient transformation matrix is:

\begin{equation*}
\mathbf{A}_{n,n} = 
\begin{pmatrix}
  1 & \left( \log\left( \frac{\nu}{\nu_a}\right) \right) & \left( \log\left( \frac{\nu}{\nu_a}\right) \right)^2\\
  0 & 1 & 2\left( \log\left( \frac{\nu}{\nu_a}\right) \right)\\
  0 & 0 & 1
\end{pmatrix}
\end{equation*}

Substituting in the values:

\begin{equation*}
\begin{pmatrix}
  1 & \left( \log\left( \frac{\nu}{\nu_a}\right) \right) & \left( \log\left( \frac{\nu}{\nu_a}\right) \right)^2\\
  0 & 1 & 2\left( \log\left( \frac{\nu}{\nu_a}\right) \right)\\
  0 & 0 & 1
\end{pmatrix}
\begin{pmatrix}
    a_0 \\
    a_1 \\
    a_2 \\
\end{pmatrix}
= \begin{pmatrix}
    b_0 \\
    b_1 \\
    b_2 \\
\end{pmatrix}
\end{equation*}

\begin{equation*}
    \begin{pmatrix}
    a_0 + a_1 \left( \log\left( \frac{\nu}{\nu_a}\right) \right) + a_2 \left( \log\left( \frac{\nu}{\nu_a}\right) \right)^2 \\
    a_0 + 2a_1\left( \log\left( \frac{\nu}{\nu_a}\right) \right) \\
    a_2 \\
\end{pmatrix}
= \begin{pmatrix}
    b_0 \\
    b_1 \\
    b_2 \\
\end{pmatrix}
\end{equation*}

\end{appendix}
\bibliographystyle{pasa-mnras}
\bibliography{paper}

\end{document}